\documentclass[pra,twocolumn,eqsecnum,superscriptaddress,showpacs,amsmath,amstex,amssymb,citeautoscript]{revtex4-1}

\usepackage[T1]{fontenc}
\usepackage[utf8]{inputenc}
\usepackage{lipsum}
\usepackage{amsmath}
\usepackage{amssymb}
\usepackage{bbm,bm}
\usepackage{braket}
\usepackage{xcolor}
\usepackage{pifont}
\newcommand{\cmark}{\ding{51}}%
\newcommand{\xmark}{\ding{55}}%
\usepackage[mathscr]{euscript}
\usepackage[shortlabels]{enumitem}
\usepackage[justification=justified]{subcaption}
\usepackage{graphicx}
\usepackage{lipsum}
\allowdisplaybreaks
\usepackage{float}
\usepackage{graphicx}
\usepackage{dsfont}
\usepackage{comment}
\usepackage[colorlinks=true]{hyperref}  
\hypersetup{
    bookmarks=true,         
    unicode=false,          
    pdftoolbar=true,        
    pdfmenubar=true,        
    pdffitwindow=false,     
    pdfstartview={FitH},    
    pdftitle={Displacement-field-tunable superconductivity \\ in an inversion-symmetric twisted van der Waals heterostructure},    
    pdfauthor={Scammell and Scheurer},     
    pdfsubject={},   
    pdfcreator={},   
    pdfproducer={}, 
    pdfkeywords={} {} {}, 
    pdfnewwindow=true,      
    colorlinks=true,       
    linkcolor=blue, 
    citecolor=blue,        
    filecolor=magenta,      
    urlcolor=blue           
}

\newcommand{\equref}[1]{Eq.~(\ref{#1})}
\newcommand{\equsref}[2]{Eqs.~(\ref{#1}) and (\ref{#2})}
\newcommand{\secref}[1]{Sec.~\ref{#1}}
\newcommand{\figref}[1]{Fig.~\ref{#1}}
\newcommand{\refcite}[1]{Ref.~\onlinecite{#1}} 
\newcommand{\refscite}[1]{Refs.~\onlinecite{#1}}

\newcommand{\tableref}[1]{Table~\ref{#1}}
\newcommand{\appref}[1]{Appendix~\ref{#1}}

\newcommand{\pdagger}{{\phantom{\dagger}}}
\newcommand{\diff}{\mathrm{d}}

\renewcommand{\approx}{\simeq}

\renewcommand{\vec}[1]{\boldsymbol{#1}}

\definecolor{wrongultramarine}{rgb}{1,0.5,0}

\begin{document}

\title{Displacement-field-tunable superconductivity \\ in an inversion-symmetric twisted van der Waals heterostructure}

\author{Harley D.~Scammell}
\affiliation{School of Mathematical and Physical Sciences, University of Technology Sydney, Ultimo, NSW 2007, Australia.}

\author{Mathias S.~Scheurer}
\affiliation{Institute for Theoretical Physics III, University of Stuttgart, 70550 Stuttgart, Germany}

\begin{abstract}
We investigate the superconducting properties of inversion-symmetric twisted trilayer graphene by considering different parent states, including spin-singlet, triplet, and SO(4) degenerate states, with or without nodal points. By placing transition metal dichalcogenide layers above and below twisted trilayer graphene, spin-orbit coupling is induced in TTLG and, due to inversion symmetry, the spin-orbit coupling does not spin-split the bands. The application of a displacement field ($D_0$) breaks the inversion symmetry and creates spin-splitting. We analyze the evolution of the superconducting order parameters in response to the combined spin-orbit coupling and $D_0$-induced spin-splitting. Utilizing symmetry analysis combined with both a direct numerical evaluation and a complementary analytical study of the gap equation, we provide a comprehensive understanding of the influence of spin-orbit coupling and $D_0$ on superconductivity. These results contribute to a better understanding of the superconducting order in twisted trilayer graphene.
\end{abstract}

\maketitle

\section{Introduction}

In recent years, graphene moir\'e superlattices with small twist angles (around 1-2 degrees) \cite{macdonald2019bilayer,andrei2020graphene}  have been studied for their potential to host various quantum many-body phases  \cite{Cao2018_correlated, Cao2018_superconductivity,Yankowitz2019_TBG,Lu2019_TBG,Choi2019, Kerelsky2019, Xie2019, Jiang2019,Sharpe2019, Serlin2020,doi:10.1126/science.abc2836,FCIExp,MoireNematic,TMDNadjPerge,TMDNadjPerge_2,doi:10.1126/science.abh2889,Park2021_tTLG, Hao_2021,PauliLimitViolation,2021arXiv210912127K,2021arXiv210912631T,JiaTrilayer,2021arXiv211207127S,Jiang-Xiazi_diode,2021arXiv211001067D,2023arXiv230212274S,2022NatPh..18..633J,2022arXiv220608354M}. However, graphene's weak intrinsic spin-orbit coupling (SOC) \cite{PhysRevB.74.165310,PhysRevB.74.155426} limits the phenomenology and, thus, the potential applications of moiré superlattices built exclusively from graphene layers. 
Enhancing SOC can unlock many additional opportunities such as stabilizing topological phases \cite{RevModPhys.82.3045,ZaletelSOC}, affecting the competition between instabilities \cite{PhysRevLett.122.246401, PhysRevLett.122.246402, XieMacDonald2020,PhysRevX.10.031034,PhysRevB.103.205414,OurPRXTTLG,2022arXiv220405317W, DiodeTheory}, and enabling spintronics applications \cite{ReviewvdWs,AnotherReviewTwistronics,PhysRevB.106.L081406}. Transition metal dichalcogenide (TMD) layers, e.g. WSe$_2$ or MoSe$_2$, can be used to induce SOC in graphene \cite{IndSOCExp1,IndSOCExp2,IndSOCExp3};  the form of the proximitized SOC terms is well established for both single-layer \cite{Gmitra2015,PhysRevB.99.075438,PhysRevB.100.085412,PhysRevB.104.195156,2022arXiv220609478L,PhysRevResearch.4.L022049}, and non-twisted multi-layer \cite{TMD_BilayerGrapheneExp, 2019arXiv190101294Z, PhysRevB.104.075126, PhysRevB.105.115126}, graphene. Notably, the resulting SOC terms induced by the TMD layer can be tuned based on the choice of TMD and the twist angle relative to graphene \cite{PhysRevB.104.195156,PhysRevB.99.075438,PhysRevB.100.085412,2022arXiv220609478L,PhysRevResearch.4.L022049,PhysRevB.104.075126,PhysRevB.105.115126}. While some experiments, e.g.~\refscite{2021arXiv211207127S,doi:10.1126/science.abh2889}, have demonstrated the impact of TMD layers on the correlated physics of graphene moiré systems, the role of the proximitized TMD layer in the observed phases is not always clear; the effect of SOC on the correlated physics of graphene moir\'e systems thereby remains an open question. 

Arguably, the role of SOC has not been elucidated due to the absence of a systematic approach to switch on/off or tune the strength of spin-orbit coupling. In a recent work \cite{MobiusPRL}, we proposed a method to achieve this. Meanwhile, in the present work, we provide a detailed analysis of the consequences for superconductivity. In particular, this work provides a detailed classification and understanding of the superconducting states of twisted trilayer graphene (TTLG) and their evolution under inversion-symmetric-proximitized SOC combined with an applied displacement field ($D_0$). The inversion-symmetry-preserving proximitization of SOC allows for spin-splitting of the electronic bands, including those comprising the superconducting states, to be directly switched on/off and tuned with applied displacement field \cite{MobiusPRL}. In this way, the displacement field has a direct influence on the superconducting states, allowing to control the mixture of different pairing channels and drive superconducting phase transitions. 
To comprehensively identify the key features of the superconducting states, we employ three complementary approaches---(i) an unbiased symmetry analysis, (ii) numerically solving the mean-field gap equations, allowing for arbitrary admixtures and momentum dependencies, and (iii) a perturbative analytical study of the gap equations; all three perspectives yield consistent results and allow us to draw a detailed picture of the evolution of superconductivity in spin-orbit-coupled TTLG. 
Moreover, it was established in \cite{MobiusPRL} that the Fermi surfaces of TTLG can exhibit a M\"obius-like spin-texture; in this work we show, analytically, why this M\"obius-texture is inherited by the superconducting order.

The rest of this paper is organized as follows: Section~\ref{ContinuumModel} establishes the continuum models and subsequent band structure for the van der Waals heterostructure system both with/without inversion-symmetry. Thereafter we focus on the inversion symmetric case. Section \ref{s:SC_model} details the mean-field/gap equation for the various possible superconducting order parameters. Section \ref{s:symm_analysis} performs a symmetry analysis of the evolution of the mean-field superconducting order parameter under combined SOC -- including Rashba and Ising -- as well as $D_0$. Section \ref{NumericalResultsGapped} performs direct numerical computations of the evolution of the eigenvalues/vectors; Sec. \ref{s:PT} examines the nature of the superconducting-to-superconducting phase transition observed in the numerics, while \secref{SC_minimal} provides analytic arguments to further elucidate key findings of our numerics. The results of the proceeding sections related to nodeless superconducting order. Section \ref{s:SC_nodal} provides an analysis of nodal pairing, which is motivated by recent experiment \cite{NodalYazdani,ExperimentNodes}. Finally, a discussion and outlook can be found in \secref{s:Discussion}.

\begin{figure}[tb]
   \centering
\includegraphics[width=1.0\linewidth]
{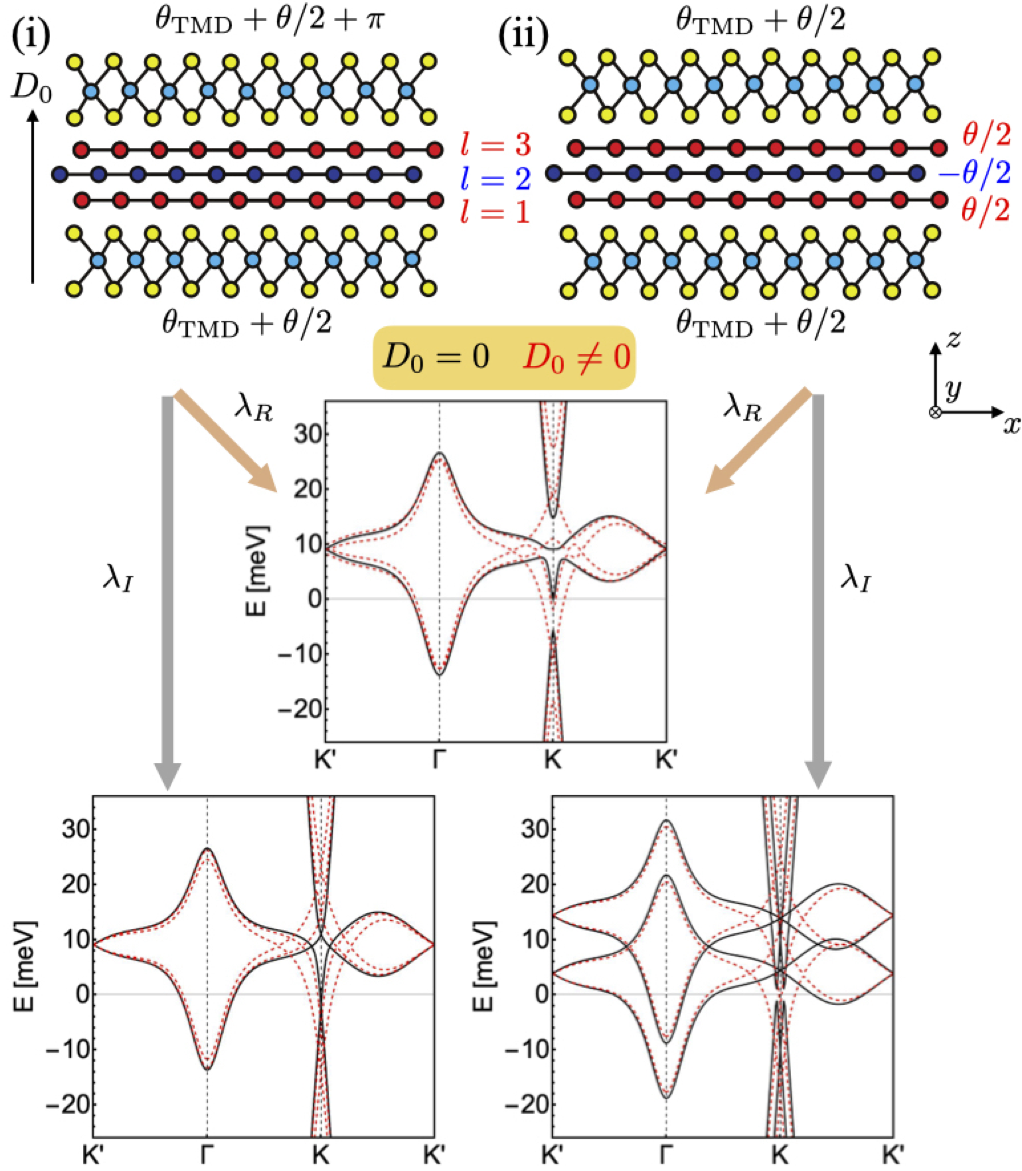}
    \caption{Two distinct classes of configurations for the TMD/TTLG/TMD heterostructure and their representative band structures. (i) Inversion symmetric heterostructure (at $D_0=0$), whereby the upper TMD layer is antiparallel to the lower layer, and (ii) mirror symmetric heterostructure (at $D_0=0$) with the TMD layers parallel. (Middle panel) Band structure for both configurations for $\{\lambda_R,\lambda_I\}=\{10,0\}$ meV and $D_0=0$ ($D_0=30$) meV in black (dashed red); band structure is identical for both configurations as long as $\lambda_I=0$. Bottom left and right:  Band structure for, respectively, the (i) and (ii) configurations with $\{\lambda_R,\lambda_I\}=\{0,10\}$ meV and $D_0=0$ ($D_0=30$) meV in black (dashed red).}
    \label{f:Bandstr}
\end{figure} 

\section{Model and symmetries}

\subsection{Continuum model}\label{ContinuumModel}

We utilize a three-layer expansion of the continuum model for twisted-bilayer graphene \cite{dos2007graphene,bistritzer2011moire,dos2012continuum,KhalafKruchkov2019,2019arXiv190712338L,PhysRevB.103.195411} to capture the band structure of the system. In order to formulate the Hamiltonian, we define $a_{\vec{k};\rho,l,\eta,s,\vec{G}}$ as the annihilation operator for an electron in graphene layer $l=1,2,3$, with momentum $\vec{k}$ in the moir\'e Brillouin zone (MBZ), and in sublattice $\rho=A,B$, valley $\eta = \pm$, spin $s=\uparrow,\downarrow$, and reciprocal moir\'e lattice (RML) vector $\vec{G} = \sum_{j} n_j \vec{G}_j$, where $n_j \in \mathbbm{Z}$. Throughout our work, we use the same notation for Pauli matrices and the corresponding quantum numbers, so that $\rho_j$, $s_j$, and $\eta_j$ are the Pauli matrices for sublattice, spin, and valley space, respectively, with $j=0,1,2,3$. To reveal the mirror symmetry $\sigma_h$ of the system, we switch to its eigenbasis \cite{KhalafKruchkov2019} by introducing the field operators $b_{\vec{k};\rho,\ell,\eta,s,\vec{G}}$,
\begin{equation}
    a_{\vec{k};\rho,l,\eta,s,\vec{G}} = V_{l,\ell} b_{\vec{k};\rho,\ell,\eta,s,\vec{G}}, \quad V=\frac{1}{\sqrt{2}} \begin{pmatrix} 1 & 0 & -1 \\ 0 & \sqrt{2} & 0 \\ 1 & 0 & 1 \end{pmatrix}, \label{TrafoToMirrorEigenbasis}
\end{equation}
with $\ell = 1,2$ ($\ell = 3$) corresponding to the mirror-even (mirror-odd) subspaces. 

The continuum Hamiltonian is divided into four distinct parts, denoted as $h_{\vec{k},\eta} = h^{(g)}_{\vec{k},\eta} + h^{(t)}_{\vec{k},\eta} + h^{(D)}_{\vec{k}} + h^{(\text{SOC})}_{\vec{k},\eta}$. These correspond to the contribution of each individual graphene layer, the tunneling between the layers, the coupling to the electric displacement field, and the SOC terms induced by proximity.  The TTLG contribution $h^{(g)}_{\vec{k},\eta} + h^{(t)}_{\vec{k},\eta}$ separates into an effective TBG, mirror-even $\ell=1,2$ subspace (denoted $h^{(\text{TBG})}_{\vec{k},\eta}$) and graphene, mirror-odd $\ell=3$ subspace (denoted $h^{(\text{G})}_{\vec{k},\eta}$), such that
\begin{align}
    \notag &\left(h^{(g)}_{\vec{k},\eta} + h^{(t)}_{\vec{k},\eta}\right)_{\rho,\ell,s,\vec{G};\rho',\ell',s',\vec{G}'}\\
    &= 
\left(\begin{array}{@{}c|c@{}}
  \begin{matrix}
  \vspace{-0.5cm} (h^{(\text{TBG})}_{\vec{k},\eta})_{\rho,s,\vec{G};\rho',s',\vec{G}'} \
  \end{matrix}
  & 0 \\ & 0 \\
\hline
  0 \ \ \ 0 &
  \begin{matrix}
  \vspace{-0.05cm}(h^{(\text{G})}_{\vec{k},\eta})_{\rho,s,\vec{G};\rho',s',\vec{G}'}
  \end{matrix}
\end{array}\right)_{\ell,\ell'}\ .
\end{align}
The explicit forms of $h^{(g)}_{\vec{k},\eta}$ and $h^{(t)}_{\vec{k},\eta}$ (or equivalently $h^{(\text{TBG})}_{\vec{k},\eta}$ and $h^{(\text{G})}_{\vec{k},\eta}$) are presented in Appendix \ref{A:Bandstr}. The displacement field mixes the mirror-even and odd subspaces, 
\begin{align}
    \notag &\left(h^{(D)}_{\vec{k}}\right)_{\rho,\ell,s,\vec{G};\rho',\ell',s',\vec{G}'}
    \\ \hspace{1cm} &= -D_0 \delta_{\rho,\rho'}\delta_{s,s'}\delta_{\vec{G},\vec{G}'} \begin{pmatrix} 0 & 0 & 1 \\ 0 & 0 & 0 \\ 1 & 0 & 0 \end{pmatrix}_{\ell,\ell'}, \label{MatrixFormOfD}
\end{align}
as do the antisymmetric terms of SOC,
\begin{align}
    \notag &\left(h^{(\text{SOC})}_{\vec{k},\eta}\right)_{\rho,\ell,s,\vec{G};\rho',\ell',s',\vec{G}'}
    \\ \hspace{1cm} &= \delta_{\vec{G},\vec{G}'} \begin{pmatrix} (h_\eta^{\text{s}})_{\rho,s;\rho',s'} & 0 & (h_\eta^{\text{a}})_{\rho,s;\rho',s'} \\ 0 & 0 & 0 \\ (h_\eta^{\text{a}})_{\rho,s;\rho',s'} & 0 & (h_\eta^{\text{s}})_{\rho,s;\rho',s'} \end{pmatrix}_{\ell,\ell'},
\end{align}
i.e.~the terms $(h_\eta^{\text{a}})_{\rho,s;\rho',s'}$. The explicit form of $h^{(\text{SOC})}$ depends on the configuration of the TMD layers; we consider the two distinct configurations shown in \figref{f:Bandstr}, which correspond to (i) inversion symmetric and (ii) mirror symmetric. Explicitly, the setup (ii) is obtained from (i) by rotating the upper TMD layer by $C_{2z}$. We consider only `Rashba' and `Ising'-type SOC, with  couplings $\lambda_{\text{R}}$ and $\lambda_{\text{I}}$, since these are known to be more dominant than any other type  \cite{PhysRevB.104.195156,PhysRevB.99.075438,PhysRevB.100.085412,2022arXiv220609478L,PhysRevResearch.4.L022049}. It is noteworthy that the relative strength of $\lambda_\text{R}$ and $\lambda_\text{I}$ can be tuned with the twist angle $\theta_{\text{TMD}}$ between the graphene and the TMD layers \cite{PhysRevB.104.195156,PhysRevB.99.075438,PhysRevB.100.085412,2022arXiv220609478L,PhysRevResearch.4.L022049,PhysRevB.104.075126,PhysRevB.105.115126}. 
Upon transforming to the mirror eigenbasis, the $h_\eta^{\text{s}}$ and $h_\eta^{\text{a}}$ contributions for the setups (i) and (ii) are found to be \begin{align}
 \notag  \text{(i)}: \ \  h_\eta^{\text{s}} &=  0, \quad
     h_\eta^{\text{a}} = -\lambda_{\text{I}} s_z \eta - \lambda_{\text{R}} \left(\eta \rho_x s_y - \rho_y s_x \right),\\    \text{(ii)}: \ \ \tilde{h}_\eta^{\text{s}} & = -\lambda_{\text{I}} s_z \eta, \quad
     \tilde{h}_\eta^{\text{a}} = - \lambda_{\text{R}} \left(\eta \rho_x s_y - \rho_y s_x \right). \label{SpinPartofHam}
\end{align}
Both the Rashba and Ising terms are odd under $C_{2z}$ and, since inversion $I = \sigma_h C_{2z}$, the $h_\eta^{\text{a}}$ ($h_\eta^{\text{s}}$) are even (odd) under inversion, c.f. Table \ref{ActionOfSymmetries}. Spin-splitting occurs when inversion symmetry is broken; for setup (i) having that $\tilde{h}_\eta^{\text{s}}=0$ means that inversion symmetry is intact and there is no spin-splitting unless $D_0\neq0$. On the other hand, for setup (ii) having $\tilde{h}_\eta^{\text{s}}\neq0$ explicitly breaks inversion-symmetry and thereby allows for spin-splitting of the bands even at $D_0=0$.
\figref{f:Bandstr} provides a comparison of the band structures for the two configurations. In line with \equref{SpinPartofHam}, we see that the two setups have identical spectrum as long as $\lambda_{\text{I}}=0$.

\begin{figure}[tb]
   \centering
    \includegraphics[width=1.0\linewidth]
    {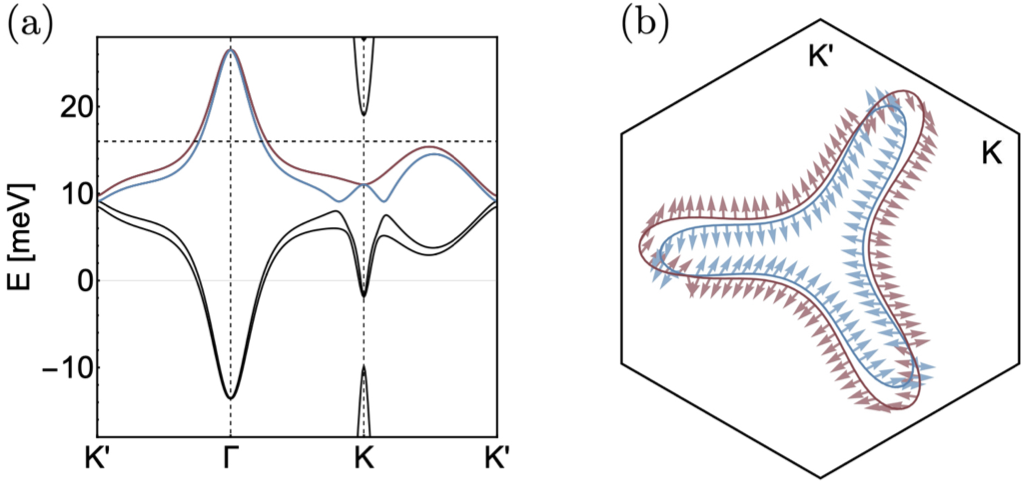}
    \caption{(a) Band structure and (b) Fermi surfaces and their spin texture (arrows). Here twist angle $\theta=1.75^\circ$, $\varepsilon_F=16$ meV, $\lambda_{\text{R}}= 20\,\textrm{meV}$, $\lambda_{\text{I}}=0$, $D_0=10$ meV and valley $\eta = +$.
    }
    \label{fig:BS_Spin}
\end{figure}

The spin-splitting of the bands under SOC and $D_0$ is captured via $\vec{g}_{\vec{k}} \neq 0$ in the effective Hamiltonian, 
\begin{align}
    h^{\text{eff}}_{\vec{k},\eta} = s_0 \xi_{\eta\cdot\vec{k}} + \eta \,\vec{g}_{\eta\cdot\vec{k}}\cdot\vec{s}\ , \label{FirstEffectiveHamiltonian}
\end{align}
for the bands of tTLG near the Fermi level, where $\xi_{\vec{k}}$ is the spin-independent part of the band structure. To be concrete, \figref{fig:BS_Spin} presents the band structure, Fermi surfaces and spin texture for the case $\lambda_R, D_0\neq0$. As discussed in detail in \refcite{MobiusPRL}, $\vec{g}_{\eta\vec k}$ exhibits vortices at three generic momenta inside the Brillouin zone which lead to three (almost) crossing points of the Fermi surface for a (close to a) specific value of the chemical potential. In \figref{fig:BS_Spin}(b), we tuned the system close to this point, which leads to the M\"obius-like nature of the Bloch spin textures. In \secref{SC_minimal} we will use the notion of the $\vec{g}_{\eta\vec k}$-vector to make analytic statements about the properties of the superconducting states.

\subsection{Superconductivity}\label{s:SC_model}
We present a minimal interacting model used to capture the various possibilities of superconductivity in TTLG, which we refer to as the {\it parent superconducting state}. For the parent superconducting state, we exclusively consider intervalley pairing, not intravalley pairing, which is expected to be dominant; this is due to the guarantee of a nesting logarithm through time-reversal symmetry.
On top of this, we account for the possibilities \cite{XuBalents2018, You2019, OurClassification, LakeSenthil} of spin-singlet dominant, spin-triplet dominant, and high-symmetry spin-SO(4) pairing, as well as both nodal and nodeless superconductivity --- corresponding to the 1D and 2D IRs of $C_{3z}^s$, respectively. 
We assume that the Cooper pairs of the parent state are formed from the partially-filled, spin-degenerate band, consistent with experimental observations \cite{Park2021_tTLG,HaoKim_tTLG,Jiang-Xiazi_diode,2021arXiv211207127S}.

\begin{figure}[t!]
\includegraphics[width=0.25\textwidth,clip]{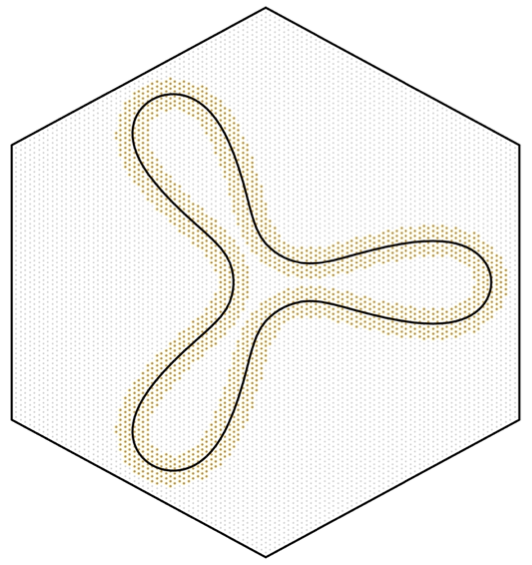}
 \caption{The discrete grid realization of the $\Theta_{\vec k;\varepsilon_F}$-functions of \equref{Gamma_eqns}; black curve is the Fermi surface of the unperturbed system at $\varepsilon_F=9.5$ meV and $\theta=1.50^\circ$, orange points indicate $\Theta_{\vec k;\varepsilon_F}=1$, gray points indicate  $\Theta_{\vec k;\varepsilon_F}=0$. 
 }
\label{fig:FSgrid}
\end{figure}

Let $\varepsilon^0_{\eta,s,\vec k}$ and $\psi^0_{\eta,s,\vec k}$ denote the spin-degenerate bands and corresponding eigenfunctions at the Fermi level of the unperturbed system $(h^{(g)}_{\eta}+h^{(t)}_{\eta})$ with quantum numbers: spin $s$; graphene valley $\eta$; and quasimomenta $\vec k$ restricted to the MBZ. We denote the corresponding creation operator $c^\dag_{\vec k,\eta,s}$. The combined perturbations of SOC and $D_0$ are captured in the perturbed eigenvectors of the noninteracting Hamiltonian, $h_{\vec{k},\eta}\psi_{\eta,n,\vec k} = \varepsilon_{\eta,n,\vec k}\psi_{\eta,n,\vec k}$, with  $h_\eta=h^{(g)}_{\eta}+h^{(t)}_{\eta} + h^{(\text{SOC})}_\eta+h^{(\text{D})}_\eta$. The band index $n$ replaces spin $s$ which is no longer a good quantum number. The electron creation operators for the Bloch states of the perturbed system, $\tilde{c}_{\vec k,\eta,n}^\dag$, are related to the unperturbed creation operators via
\begin{align}
\label{overlap}
    c^\dag_{\vec k,\eta,s}=\sum_n C^*_{\eta, n, s, \vec k} \tilde{c}^\dag_{\vec k,\eta,n}, \, C^*_{\eta, n, s, \vec k}\equiv \braket{\psi_{\eta,n,\vec k} | \psi^0_{\eta,s,\vec k}}. 
\end{align}
 The mean-field Hamiltonian, decoupled into the Cooper channel for intervalley pairing, is valley-diagonal ${\cal H} = \sum_{\eta=\pm} {\cal H}_\eta $ with
  \begin{align}
  \label{supp_HBdG}
   &{\cal H}_\eta =\sum_{\vec k, n} \varepsilon_{\eta,n,\vec k} \tilde{c}_{\vec k,\eta,n}^\dag \tilde{c}^\pdagger_{\vec k,\eta,n}\\
 \notag &+ \sum_{\vec k_1, \vec k_2;\mu,\nu} (-\Gamma^{-1})_{\vec k_1, \vec k_2; \mu,\nu} \bigl(d_{\vec k_1,\eta}^*\bigr)_\mu \bigl(d_{\vec k_2,\eta}\bigr)_\nu \\
 \notag  &+ \sum_{\vec k}\sum_{n,n'}\sum_{s_1,s_2}\sum_{\mu}\Big\{ \tilde{c}_{\vec k, \eta,n}^\dag \tilde{c}_{-\vec k,\eta,n'}^\dag\\
 \notag  & \quad \times \left[(d_{\vec k,\eta})_\mu (s_\mu i s_y )_{s_1,s_2} C^*_{\eta,n,s_1,\vec k} C_{\eta,n',s_2,\vec k}\right]   +  \text{H.c.} \Big\}.
  \end{align}
  Here the intervalley superconducting order parameter $d_{\mu,\vec k, \eta}$ encodes the spin, quasi-momentum and valley structure, where $\mu=0$ refers to spin-singlet and $\mu=1,2,3$ refer to the components of the spin-triplet.
  We point out that the mean-field description of the {\it parent} superconducting state corresponds to the limit $C_{\eta, n, s, \vec k}=1$; for the {\it perturbed} superconducting state, the overlap factors $C_{\eta, n, s, \vec k}$ encode the SOC-induced mixing of the $\mu$-components of $d_{\mu,\vec k, \eta}$ as a function of $\vec k$.
  Here we allow the vertex $\Gamma_{\vec k_1, \vec k_2; \mu,\nu}$ to be either even or odd under: $\vec{k}_1\to -\vec{k}_1$ or $\vec{k}_2\to-\vec{k}_2$. The even (odd) vertex is introduced to accommodate nodeless (nodal) superconductivity. Explicitly, $\Gamma_{\vec k_1, \vec k_2; \mu,\nu}=\Gamma^\text{even}_{\vec k_1, \vec k_2; \mu,\nu}$ or  $\Gamma^\text{odd}_{\vec k_1, \vec k_2; \mu,\nu}$, with
 \begin{align}
 \label{Gamma_eqns}
\notag \Gamma^\text{even}_{\vec k_1, \vec k_2; \mu,\nu}&= (\gamma_0 \delta_{\mu,\nu}  + \delta\gamma \delta_{\mu,0} \delta_{\nu,0}) F_{\vec k_1, \vec k_2} \Theta_{\vec k_1;\varepsilon_F} \Theta_{\vec k_2;\varepsilon_F}\ ,\\
\Gamma^\text{odd}_{\vec k_1, \vec k_2; \mu,\nu}&=\Gamma^\text{even}_{\vec k_1, \vec k_2; \mu,\nu}\times({\cal X}_{\vec k_1} {\cal X}_{\vec k_2}+{\cal Y}_{\vec k_1} {\cal Y}_{\vec k_2}).
\end{align}
Here $F_{\vec k_1, \vec k_2}$ is a momentum dependent function (detailed in the Appendix \ref{A_symmetrizing}) while $\Theta_{\vec k;\varepsilon_F}$ is a step-function such that: $\Theta_{\vec k;\varepsilon_F}=1$ for $\vec k$ within a radius $\Lambda$ of any Fermi momentum $\vec k_F$, and $\Theta_{\vec k;\varepsilon_F}=0$ elsewhere. A representative plot of $\Theta_{\vec k;\varepsilon_F}$ is shown in \figref{fig:FSgrid}. An attractive interaction requires $\gamma_0>0$. Meanwhile, the singlet-triplet asymmetry parameter $|\delta\gamma|/\gamma_0<1$ distinguishes three cases of the parent superconducting state: (i) $\delta \gamma>0$ favors spin-singlet, (ii) $\delta \gamma<0$ favors spin-triplet, and (iii) for $\delta \gamma=0$, spin-singlet and triplet are degenerate [spin SO(4)]. Finally, in Eq. \eqref{Gamma_eqns} we have included the basis functions ${\cal X}_{\vec k},{\cal Y}_{\vec k}$, which are MBZ-periodic functions transforming as $k_x$, $k_y$ under $C_{3z}$, and are specifically taken to be
\begin{align}
\label{basis_XY}
{\cal X}_{\bm k}&=\frac{2}{\sqrt{3}} \sin \left(\frac{\sqrt{3}}{2}a_{\theta } k_x\right)
   \cos \left(\frac{1}{2}a_{\theta }
   k_y\right),\\
\notag {\cal Y}_{\bm k}&=\frac{2}{3} \left(\cos \left(\frac{\sqrt{3}}{2}
   a_{\theta } k_x\right) \sin \left(\frac{1}{2} a_{\theta
   } k_y\right)+\sin \left(a_{\theta }
   k_y\right)\right),
\end{align}
with $a_\theta$ the magnitude of the moir\'e lattice vector. The chosen combination of ${\cal X}_{\vec k}, {\cal Y}_{\vec k}$ in $\Gamma^\text{odd}_{\vec k_1, \vec k_2; \mu,\nu}$ [Eq.~\eqref{Gamma_eqns}] preserves $C_{3z}$.

 Following \eqref{supp_HBdG} we arrive at the linearized gap equation, 
\begin{align}
\label{gapeq}
&\bigl(d_{\vec k_1,\eta}\bigr)_\mu = \sum_{\nu, \vec k_2}\Gamma_{\mu,\mu',\vec k_1,\vec k_2} {\cal W}_{\mu'\nu, \vec k_2,\eta} \bigl(d_{\vec k_2,\eta}\bigr)_\nu,\\
\notag &{\cal W}_{\mu\nu,\vec k,\eta} = \sum_{n_1,n_2} \sum_{ s_1,s_2,s_3,s_4}\frac{\tanh\left(\frac{\varepsilon_{\eta,n_1\vec k}}{2T}\right)+\tanh\left(\frac{\varepsilon_{\eta,n_2\vec k}}{2T}\right)}{2(\varepsilon_{\eta,n_1\vec k}+\varepsilon_{\eta,n_2\vec k})}\\
\notag &(\sigma_\mu)_{s_2,s_3} C_{\eta,n_1,s_1,\vec k} C^*_{\eta,n_1 s_2, \vec k} C_{\eta,n_2,s_3, \vec k} C^*_{\eta,n_2,s_4,\vec k} (\sigma_\nu)_{s_4,s_1}.
\end{align}
We note that the gap equation is diagonal in $\eta$, and hence for the numerical analysis presented in Sec. \ref{s:SC_numerics} we specialize to $\eta=+1$. In fact, this uniquely determines the superconducting order parameter since its form in the other valley just follows from the Fermi Dirac constraint [see \equref{FDConstraint} below]. Further details about solving the gap equation are presented in Appendix \ref{A_symmetrizing}.

\begin{table*}[tb]
\begin{center}
\caption{Action of the point symmetries of the continuum theory on the microscopic operators $\psi_{\vec{k}}$ in the continuum model, see \secref{ContinuumModel}, and the low-energy fermions $c_{\vec{k}}$ used, e.g., in \equsref{overlap}{SuperconductingOrderParameter}. For convenience of the reader, we also list redundant symmetries. In the last two columns, we state the constraints on $D_0, \lambda_{\text{R}}$, and $\lambda_{\text{I}}$ for the respective symmetry to be present for (i) the inversion and (ii) the mirror symmetric geometry, see \equref{SpinPartofHam} and \figref{f:Bandstr}.}
\label{ActionOfSymmetries}
\begin{ruledtabular}
 \begin{tabular} {cccccc} 
Symmetry $S$ & unitary?  & $S\psi_{\vec{k};\ell,\vec{G}}S^\dagger$ & $S c_{\vec{k}}S^\dagger$ & condition for (i) & condition for (ii) \\ \hline
SO(3)$_{s}$ & \cmark    & $e^{i\vec{\varphi}\cdot \vec{s}/2}  \psi_{\vec{k};\ell,\vec{G}}$  & $e^{i\vec{\varphi}\cdot \vec{s}/2}  c_{\vec{k}}$ &  $\lambda_{\text{R}}=\lambda_{\text{I}}=0$ & $\lambda_{\text{R}}=\lambda_{\text{I}}=0$  \\ 
SO(2)$_{s}$ & \cmark    & $e^{i \varphi s_z/2}  \psi_{\vec{k};\ell,\vec{G}}$ & $e^{i\varphi s_z/2}  c_{\vec{k}}$  & $\lambda_{\text{R}}=0$ & $\lambda_{\text{R}}=0$  \\ \hline
$C_{3z}$ & \cmark    & $e^{i\frac{2\pi}{3} \rho_z\eta_z}  \psi_{C_{3z}\vec{k};\ell,C_{3z}\vec{G}}$  & $c_{C_{3z}\vec{k}}$ & $\lambda_{\text{R}}=0$ & $\lambda_{\text{R}}=0$  \\
$C^s_{3z}$ & \cmark    & $e^{i\frac{2\pi}{3} (\rho_z\eta_z + s_z)}  \psi_{C_{3z}\vec{k};\ell,C_{3z}\vec{G}}$ & $e^{i\frac{2\pi}{3} s_z}c_{C_{3z}\vec{k}}$ & --- & ---  \\ \hline
$C_{2z}$ & \cmark   & $\eta_x \rho_x \psi_{-\vec{k};\ell,-\vec{G}}$ & $\eta_x c_{-\vec{k}}$ & $\lambda_{\text{R}}=\lambda_{\text{I}}=0$ & $\lambda_{\text{R}}=\lambda_{\text{I}}=0$   \\
$C^s_{2z}$ & \cmark   & $s_z\eta_x \rho_x \psi_{-\vec{k};\ell,-\vec{G}}$ & $s_z\eta_x c_{-\vec{k}}$ & $\lambda_{\text{I}}=0$ & $\lambda_{\text{I}}=0$  \\
$C^{s'}_{2z}=C^s_{2z} i  s_{y,x}$ & \cmark   & $s_{x,y}\eta_x \rho_x \psi_{-\vec{k};\ell,-\vec{G}}$ & $s_{x,y}\eta_x c_{-\vec{k}}$ & $\lambda_{\text{R}}=0$ & $\lambda_{\text{R}}=0$  \\ \hline
$\sigma_h$ & \cmark   & $ (1,1,-1)_\ell \psi_{\vec{k};\ell,\vec{G}}$ & $\pm c_{\vec{k}}$  & $D_0=\lambda_{\text{R}}=\lambda_{\text{I}}=0$ & $D_0=\lambda_{\text{R}}=0$  \\
$\sigma^s_h$ & \cmark    & $s_z(1,1,-1)_\ell \psi_{\vec{k};\ell,\vec{G}}$ & $\pm s_z c_{\vec{k}}$  & $D_0=\lambda_{\text{I}}=0$ & $D_0=0$  \\
$\sigma^{s'}_h = \sigma^s_h i s_{y,x}$ & \cmark   & $s_{x,y}(1,1,-1)_\ell \psi_{\vec{k};\ell,\vec{G}}$ &  $\pm s_{x,y} c_{\vec{k}}$ & $D_0=\lambda_{\text{R}}=0$  & $D_0=\lambda_{\text{R}}=\lambda_{\text{I}}=0$ \\ \hline
$I = C_{2z}\sigma_h = C^s_{2z}\sigma^s_h$ & \cmark    & $ \eta_x\rho_x(1,1,-1)_\ell \psi_{-\vec{k};\ell,\vec{G}}$ & $\pm \eta_x c_{-\vec{k}}$ & $D_0=0$ & $D_0=\lambda_{\text{I}}=0$  \\ \hline
$\Theta$ & \xmark    & $\eta_x \psi_{-\vec{k};\ell,-\vec{G}}$ & $\eta_x c_{-\vec{k}}$ & $\lambda_{\text{R}}=\lambda_{\text{I}}=0$ & $\lambda_{\text{R}}=\lambda_{\text{I}}=0$  \\
$\Theta^s$ & \xmark   & $is_y\eta_x \psi_{-\vec{k};\ell,-\vec{G}}$ & $is_y\eta_x c_{-\vec{k}}$  & --- & ---  \\
 \end{tabular}
\end{ruledtabular}
\end{center}
\end{table*}

\section{Symmetry analysis}\label{s:symm_analysis}
Before solving numerically for superconductivity in the next section, we here use symmetry arguments to derive the evolution of the structure of the superconducting instabilities when $\lambda_{\text{R}}$, $\lambda_{\text{I}}$, and $D_0$ are turned on. As follows from \tableref{ActionOfSymmetries} and as summarized in \figref{fig:EvolutionofPairingAndSyms}(a), depending on which combination of these three parameters is non-zero, the system exhibits a variety of point groups. This is important for pairing since certain order parameter configurations, corresponding to distinct IRs of the point group $C_{6h} \times \text{SO}(3)_s$ of the system at $\lambda_{\text{R}}=\lambda_{\text{I}}=D_0=0$, can mix when the symmetry group is reduced.

To define the superconducting order parameter, we use the electronic operators $c^\dagger_{\vec{k},\eta,s}$ introduced in \secref{s:SC_model}, which create an electron with momentum $\vec{k}\in\text{MBZ}$ in the band in valley $\eta$ and of spin $s$ that is closest to the Fermi level in the limit $\lambda_{\text{R}}=\lambda_{\text{I}}=D_0=0$. The representations of the symmetries in this basis are listed in \tableref{ActionOfSymmetries} (which fixes their phase ambiguity). Focusing as before on the energetically most favorable intervalley pairing, the order parameter reads as
\begin{equation}
    \mathcal{H}_{\text{SC}} = \hspace{-0.3em}\sum_{\vec{k},s,s',\eta} \sum_{\mu=0}^3 c^\dagger_{\vec{k},\eta,s} \left[\left(d_{\vec{k},\eta}\right)_\mu s_\mu i s_y\right]_{s,s'} \hspace{-0.1em}c^\dagger_{-\vec{k},-\eta,s'} + \text{H.c.}, \label{SuperconductingOrderParameter}
\end{equation}
where 
\begin{equation}
    \left(d_{\vec{k},\eta}\right)_0 = \left(d_{-\vec{k},-\eta}\right)_0 \, \text{and} \,  \left(d_{\vec{k},\eta}\right)_j = -\left(d_{-\vec{k},-\eta}\right)_j, \label{FDConstraint}
\end{equation}
$j=x,y,z$, are the singlet and triplet components of the order parameter, respectively. Let us start in the limit $\lambda_{\text{R}}=\lambda_{\text{I}}=D_0=0$. As the point group $C_{6h} \times \text{SO}(3)_s$ contains spin rotation symmetry, spin-singlet and spin-triplet cannot mix. Furthermore, we expect the dominant pairing to involve Cooper pairs of electrons of the same mirror-symmetry ($\sigma_h$) sector and not between different sectors, leading to pairing being even under $\sigma_h$. In fact our low-energy description of pairing in \equref{SuperconductingOrderParameter} automatically implies that: in the presence of $\sigma_h$, every band has a distinct eigenvalue under $\sigma_h$ and the band in a given valley $\eta$ closest to the Fermi level at momentum $\vec{k}$, where an electron is created by application of $c^\dagger_{\vec{k},\eta,s}$, has the same $\sigma_h$ eigenvalue as the band closest to the Fermi level at $-\vec{k}$ and $-\eta$. 
Since the Fermi-Dirac constraint (\ref{FDConstraint}) further implies \cite{OurClassification} that singlet (triplet) is even (odd) under $C_{2z}$ and $I$, we are left with the IRs $A_g$ or $E_{2g}$ of $C_{6h}$ for singlet and $B_u$ or $E_{1u}$ for triplet pairing.

\subsection{Pairing in one-dimensional IRs of $C_{6h}$} 
\begin{figure*}[t]
   \centering
    \includegraphics[width=0.85\linewidth]
    {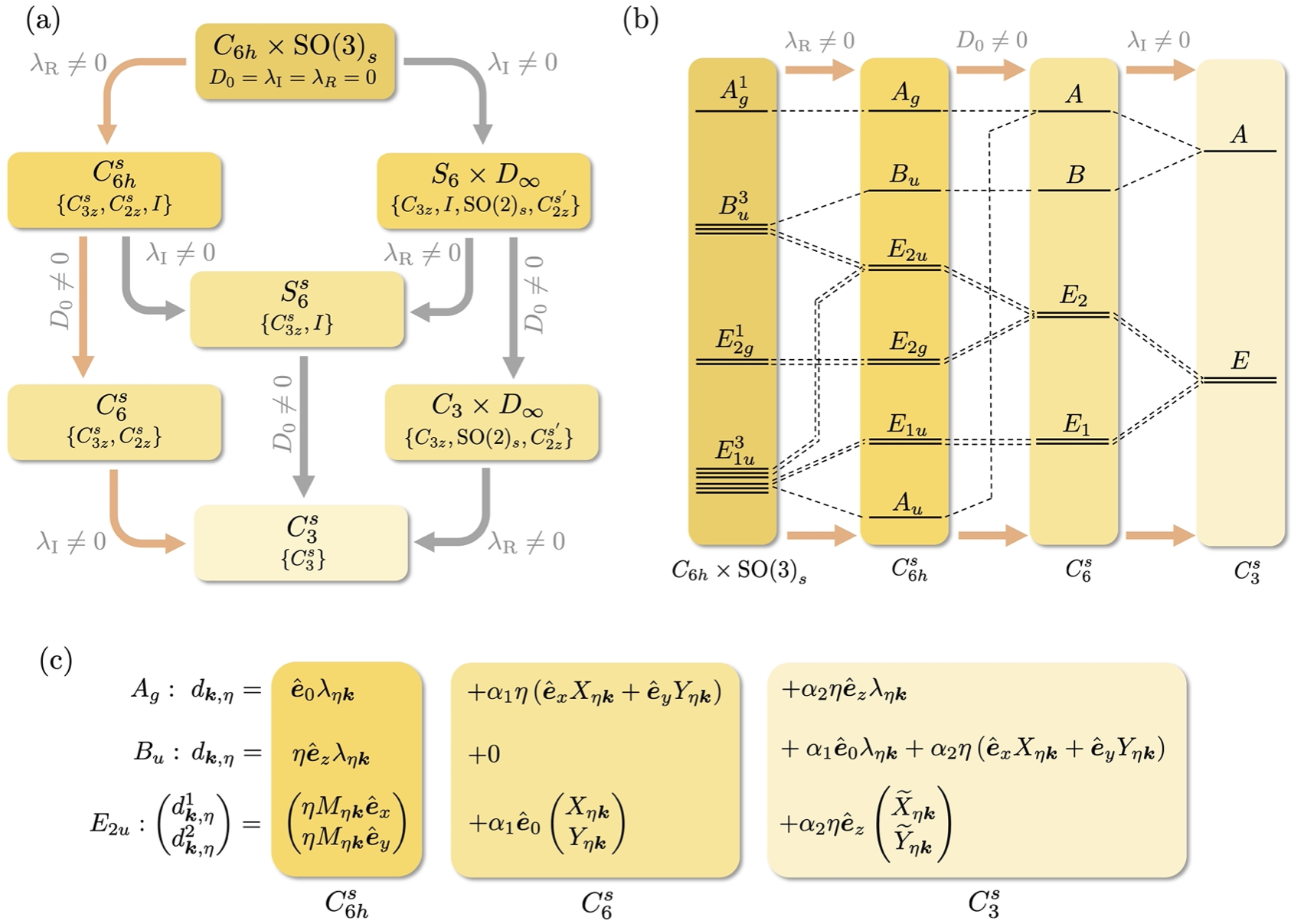}
    \caption{(a) Summary of point groups when $D_0$, $\lambda_{\text{R}}$, and $\lambda_{\text{I}}$ are turned on, focusing for concreteness on geometry (i) in \equref{SpinPartofHam}; the modifications for geometry (ii) follow from \tableref{ActionOfSymmetries}. We also list a (in many cases redundant) set of generators, see \tableref{ActionOfSymmetries} for the definitions and representations of the symmetry operations. We use $D_\infty$ to refer to the group generated by $\{\text{SO(2)}_s, C_{2z}^{s'}\}$. Choosing one specific path [indicted in light red in (a)] as an example, we show in (b) the evolution of the IRs of the pairing states. For the superconducting states emerging out of $A_g^1$ and $B_{u}^3$ we summarize their order parameters in (c). To keep the notation compact we use the four-component basis vectors $(\vec{e}_\mu)_{\mu'} = \delta_{\mu,\mu'}$.}
    \label{fig:EvolutionofPairingAndSyms}
\end{figure*}

Let us first focus on the one-dimensional IRs, $A_g$ and $B_u$, i.e., pairing transforming trivially under $C_{3z}$; to indicate their respective spin-structure, these states are represented by $A_g^1$ and $B_u^3$ in \figref{fig:EvolutionofPairingAndSyms}(b). They have the following order parameters:
\begin{align}
    A_g^1:\quad &d_{\vec{k},\eta} = (\lambda_{\eta\cdot \vec{k}};0,0,0), \\ B_u^3:\quad &d_{\vec{k},\eta} = (0;\eta \lambda_{\eta\cdot \vec{k}}\hat{\vec{n}}),
\end{align}
where $\lambda_{\vec{k}} = \lambda_{C_{3z}\vec{k}} \in \mathbb{R}$ and $\hat{\vec{n}}$ is a real unit vector (here and in the following, we employ the slight abuse of notation where the symbol $C_{3z}$ refers both to the transformation as a group element and to its vector representation). 

There are several orders of turning on the perturbations $\lambda_{\text{R}}$, $\lambda_{\text{I}}$, $D_0$, corresponding to the different paths (arrows) in \figref{fig:EvolutionofPairingAndSyms}(a). As it is most important for our discussion here, we will focus on geometry (i) in \equref{SpinPartofHam} and $\lambda_{\text{R}}$ being turned on first and then $D_0$ as an example [path in light red in \figref{fig:EvolutionofPairingAndSyms}(a)]; the generalization to other paths is straightforward. First, setting $\lambda_{\text{R}}\neq 0$ will reduce the point group to $C_{6h}^s$, where all symmetry operations in $C_{6h}$ are replaced by appropriate combinations of spatial and spin transformations, i.e., the (redundant) generators $C_{3z}$, $C_{2z}$, and $\sigma_h$ of $C_{6h}$ are replaced by their spinful counterparts $C_{3z}^s$, $C_{2z}^s$, and $\sigma_h^s$, see \tableref{ActionOfSymmetries}.
Importantly, $C_{6h}^s$ still contains $I = C_{2z}^s \sigma_h^s$ such that singlet and triplet continue to transform under different IRs and, hence, cannot mix. 
While the singlet is unaffected, the $B_u^3$ triplet splits into a state with triplet vector pinned along the $s_z$ direction (as a consequence of $C_{2z}^s$) transforming under the one-dimensional IR $B_u$ of $C_{6h}^s$ and a doublet transforming under the two-dimensional IR $E_{2u}$; their order parameters read as
\begin{align}
    B_u:\quad &d_{\vec{k},\eta} = (0;0,0,\eta \lambda_{\eta\cdot\vec{k}}), \label{BuTriplet}
    \\ E_{2u}:\quad &\begin{pmatrix} d^1_{\vec{k},\eta} \\ d^2_{\vec{k},\eta}  \end{pmatrix} = \begin{pmatrix} (0;\eta M_{\eta\cdot\vec{k}}\hat{\vec{e}}_x,0) \\ (0;\eta M_{\eta\cdot\vec{k}}\hat{\vec{e}}_y,0) \end{pmatrix}, \label{TripletPairingC6hs}
\end{align}
where $M_{\vec{k}}$ is a real, $2\times 2$ matrix, obeying $C_{3z}^{-1}M_{C_{3z}\vec{k}}C_{3z} = M_{\vec{k}}$. Naturally, for the $E_{2u}$ state emerging out of and being close to the $B_{u}^3$ (and its descendent below), we expect $M_{\vec{k}} \approx \sigma_0 \lambda_{\vec{k}}$; this is indeed what we see in the numerics (cf.~first two rows in \figref{f:vecsSO3_R}).

When also $D_0$ is non-zero, $\sigma_h^s$ is broken and the point group is reduced to $C_{6}^s$. As it does not contain inversion symmetry $I$ (nor $C_{2z}$) anymore, spin-singlet and triplet can now mix. More precisely, the presence of $C_{2z}^s$ guarantees that the triplet component admixed to the singlet only contains in-plane spin components; the $A_g$ singlet now becomes the $A$ state of $C_{6}^s$ with order parameter
\begin{equation}
    A: \quad d^A_{\vec{k},\eta} = (\lambda_{\eta\cdot \vec{k}};0,0,0) + \alpha_1 \eta (0;X_{\eta\cdot \vec{k}},Y_{\eta\cdot\vec{k}},0), \label{SingletWithTriplet}
\end{equation}
where $X_{\vec{k}}$, $Y_{\vec{k}}$ are real-valued, MBZ-periodic functions transforming as $k_x$, $k_y$ under $C_{3z}$ [such as the two components of the spin-orbit vector $\vec{g}_{\vec{k}}$ in \equref{Heff}]; we distinguish $X_{\vec{k}}$, $Y_{\vec{k}}$, which are generic functions, from the specific choice ${\cal X}_{\vec k},{\cal Y}_{\vec k}$ of \eqref{basis_XY}. Here $\alpha_1 \in \mathbb{R}$ describes the strength of the triplet admixture (coming from $A_u$ of $C_{6h}^s$). 

By the same token, $C_{2z}^s$ prevents the $B_u$ triplet in \equref{BuTriplet} to exhibit singlet or in-plane triplet components and, thus, remains of the same form; it will be relabeled as the $B$ state of $C_6^s$. Being even under $C_{2z}^s$, the $E_{2u}$ doublet in \equref{TripletPairingC6hs}, however, can mix with a singlet component (coming from $E_{2g}$ of $C_{6h}^s$) and becomes the $E_2$ state of $C_{6}^s$ with order parameter
\begin{equation}
    E_{2}:\, \begin{pmatrix} d^1_{\vec{k},\eta} \\ d^2_{\vec{k},\eta}  \end{pmatrix} = \begin{pmatrix} (0;\eta M_{\eta\vec{k}}\hat{\vec{e}}_x,0) \\ (0;\eta M_{\eta\vec{k}}\hat{\vec{e}}_y,0) \end{pmatrix} + \alpha_2 \begin{pmatrix} (X_{\eta\vec{k}};\vec{0}) \\ (Y_{\eta\vec{k}};\vec{0}) \end{pmatrix}. \label{E2State}
\end{equation}

Finally, let us also take $\lambda_{\text{I}}$ to be finite, which breaks $C_{2z}^s$ leaving us with the point group $C_{3}^s$, just consisting of spinfull three-fold rotations along the $z$ axis. Since the subduced representations of $A$ and $B$ of $C_{6}^s$ onto $C_{3}^s$ are both $A$ of $C_{3}^s$, the singlet $A$ in \equref{SingletWithTriplet} and the out-of-plane triplet in \equref{BuTriplet} hybridize and become a single phase with order parameter ($\beta \in \mathbb{R}$)
\begin{equation}
    A:\,  d_{\vec{k},\eta} = (\lambda_{\eta\vec{k}};\vec{0}) + \alpha_1 \eta (0;X_{\eta\vec{k}},Y_{\eta\vec{k}},0) + \beta (0;\eta \lambda_{\eta\vec{k}}\hat{\vec{e}}_z).
    \label{SingletRI}
\end{equation}

The broken $C_{2z}^s$ also allows for more terms in the previous $E_2$ state in \equref{E2State} which now also exhibits a spin component along the out-of-plane direction (coming from $E_1$ of $C_6^s$),
\begin{align}
    E:\quad \begin{pmatrix} d^1_{\vec{k},\eta} \\ d^2_{\vec{k},\eta}  \end{pmatrix} &= \begin{pmatrix} (0;\eta M_{\eta\vec{k}}\hat{\vec{e}}_x,0) \\ (0;\eta M_{\eta\vec{k}}\hat{\vec{e}}_y,0) \end{pmatrix} + \alpha_2 \begin{pmatrix} (X_{\eta\vec{k}};\vec{0}) \\ (Y_{\eta\vec{k}};\vec{0}) \end{pmatrix} \nonumber \\ &+ \alpha_3 \eta \begin{pmatrix} (0;\widetilde{X}_{\eta\vec{k}}\hat{\vec{e}}_z) \\ (0;\widetilde{Y}_{\eta\vec{k}}\hat{\vec{e}}_z) \end{pmatrix},
    \label{TripletRI}
\end{align}
where the tilde of $\widetilde{X}$ and $\widetilde{Y}$ in the last term just indicates that these functions need not be identical to $X$ and $Y$ but exhibit the same transformation behavior.  
The admixture of the different IRs of the respective point groups and the form of the pairing states discussed here are summarized schematically in \figref{fig:EvolutionofPairingAndSyms}(b) and (c), respectively.

\subsection{Two-dimensional IRs of $C_{6h}$}\label{SymmetryDiscussion2DIRs}
While all of the states above can and are generically expected to be fully gapped, recent experiments \cite{NodalYazdani,ExperimentNodes} indicate that also nodal pairing can be realized in graphene moir\'e systems. For this reason, we next discuss pairing emerging out of the remaining, two-dimensional IRs---$E_{2g}$ and $E_{1u}$ of $C_{6h}$. As discussed above already, they are pure singlet and triplet states, respectively, in the presence of inversion symmetry. We begin with the singlet, $E_{2g}^1$, with order parameter
\begin{equation}
    E_{2g}^1:\quad \begin{pmatrix} d^1_{\vec{k},\eta} \\ d^2_{\vec{k},\eta}  \end{pmatrix} =  \begin{pmatrix} (X_{\eta\vec{k}};0,0,0) \\ (Y_{\eta\vec{k}};0,0,0) \end{pmatrix}. \label{E2g1StateSynAn}
\end{equation}
The associated nematic superconducting state, $(d_{\vec{k},\eta})_\mu = \delta_{\mu,0}X_{\eta\vec{k}}$, can have stable nodal points (depending on the Fermi surface), while the chiral state, $(d_{\vec{k},\eta})_\mu = \delta_{\mu,0}(X_{\eta\vec{k}}+i Y_{\eta\vec{k}})$, will generically be fully gapped. Finite $\lambda_{\text{R}}$ does not change the form of the order parameter since $I$ (of $C_{6h}^s$) prohibits any triplet admixture. However, when also $D_0$ is finite, the $E_{2g}$ and $E_{2u}$ states of $C_{6h}^s$ can hybridize into the $E_2$ state of $C_6^s$,
\begin{equation}
    \begin{pmatrix} d^1_{\vec{k},\eta} \\ d^2_{\vec{k},\eta}  \end{pmatrix} = \begin{pmatrix} (X_{\eta\vec{k}};0,0,0) \\ (Y_{\eta\vec{k}};0,0,0) \end{pmatrix} + \alpha_1 \begin{pmatrix} (0;\eta M_{\eta\vec{k}}\hat{\vec{e}}_x,0) \\ (0;\eta M_{\eta\vec{k}}\hat{\vec{e}}_y,0) \end{pmatrix}. \label{GappingOutAdmixture}
\end{equation}
By design, the form of this state is equivalent to that in \equref{E2State} discussed above. However, there are a few important differences in the precise nature of it: if the $E_{2g}^1$ state dominates at $\lambda_{\text{R}}=\lambda_{\text{I}}=D_0=0$, then turning on $\lambda_{\text{R}}, D_0$ weakly will lead to a small admixture of the second term in \equref{GappingOutAdmixture} to the first. Furthermore, as opposed to \equref{E2State}, there is no reason anymore that $M_{\eta\vec{k}}$ in \equref{GappingOutAdmixture} is close to $\sigma_0 \lambda_{\vec{k}}$ with $\lambda_{\vec{k}}$ that is approximately constant on the Fermi surface. And, indeed, we find a non-trivial $M_{\vec{k}}$ in our numerics below.  

Finally, the triplet $E_{1u}^3$ has a matrix-valued order parameter,
\begin{equation}
\label{E1u3}
    d^{i,j}_{\vec{k},\eta} = \eta \begin{pmatrix} X_{\eta\vec{k}} \\ Y_{\eta\vec{k}}  \end{pmatrix}_{\hspace{-0.4em}j} \hat{\vec{e}}_i , \quad i=x,y,z, \,\,j=1,2,
\end{equation}
which leads to a multitude of possible phases in the absence of SOC \cite{OurClassification}. When $\lambda_{\text{R}}$ is turned on, it splits into 
\begin{align}
    &A_u:\quad d_{\vec{k},\eta} = \eta(0;X_{\eta\vec{k}},Y_{\eta\vec{k}},0), \label{AuState} \\ 
    \label{E1u}
    &E_{1u}:\quad \begin{pmatrix} d^1_{\vec{k},\eta} \\ d^2_{\vec{k},\eta}  \end{pmatrix} = \eta\begin{pmatrix} (0;X_{\eta\vec{k}}\hat{\vec{e}}_z) \\ (0;Y_{\eta\vec{k}}\hat{\vec{e}}_z) \end{pmatrix}
\end{align}
and another $E_{2u}$ component that mixes with the $B_u^3$ state [absorbed in $M_{\vec{k}}$ in \equref{TripletPairingC6hs}]. 
If also $D_0$ is finite, the $A_u$ state will mix with the $A_g$ state leading to the $A$ superconducting order parameter of $C_6^s$ in \equref{SingletWithTriplet}. Interestingly, the $E_{1u}$ state of $C_{6h}^s$ simply becomes the $E_1$ state of $C_{6}^s$ without any changes to its form. Finally, for finite $\lambda_{\text{I}}$, the $E_{1}$ and $E_{2}$ phases of $C_6^s$ mix and become the $E$ state of $C_{3}^s$ with order parameter of the form of \equref{TripletRI}.

\section{Superconducting energetics}\label{s:SC_numerics}

Our next objective is to explicitly compute the influence of $D_0$ and SOC on the various possible parent states. The numerical results will be directly related back to the symmetry classification of the previous section, and we will further appeal to analytic arguments, Sec \ref{SC_minimal}, to explain key features of our numerical findings.

\subsection{Numerics for fully gapped states}\label{NumericalResultsGapped}

\begin{figure}[t]
   \centering
    \includegraphics[width=\linewidth]
    {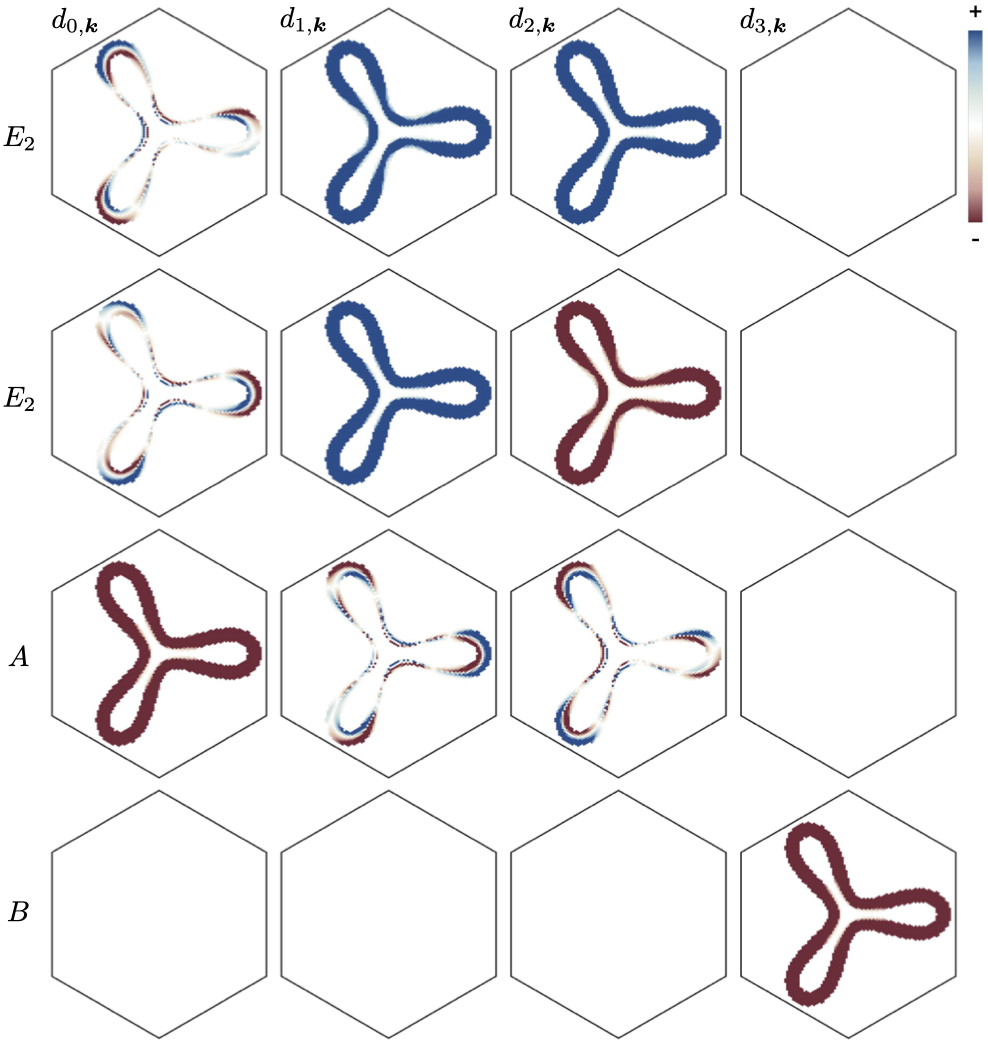}
    \caption{Spin-triplet favored parent state, i.e.  $\delta\gamma/\gamma_0=-0.25$, with fixed $D_0=6$ meV, $\lambda_{\text{R}}=30$meV, $\lambda_{\text{I}}=0$ and $\varepsilon_F=9.6$ meV.  (Top-to-bottom) Corresponds to leading eigenvalue down to fourth eigenvalue; the corresponding IRs are labeled.}
    \label{f:vecsSO3_R}
\end{figure}

\begin{figure}[t]
   \centering
    \includegraphics[width=\linewidth]
    {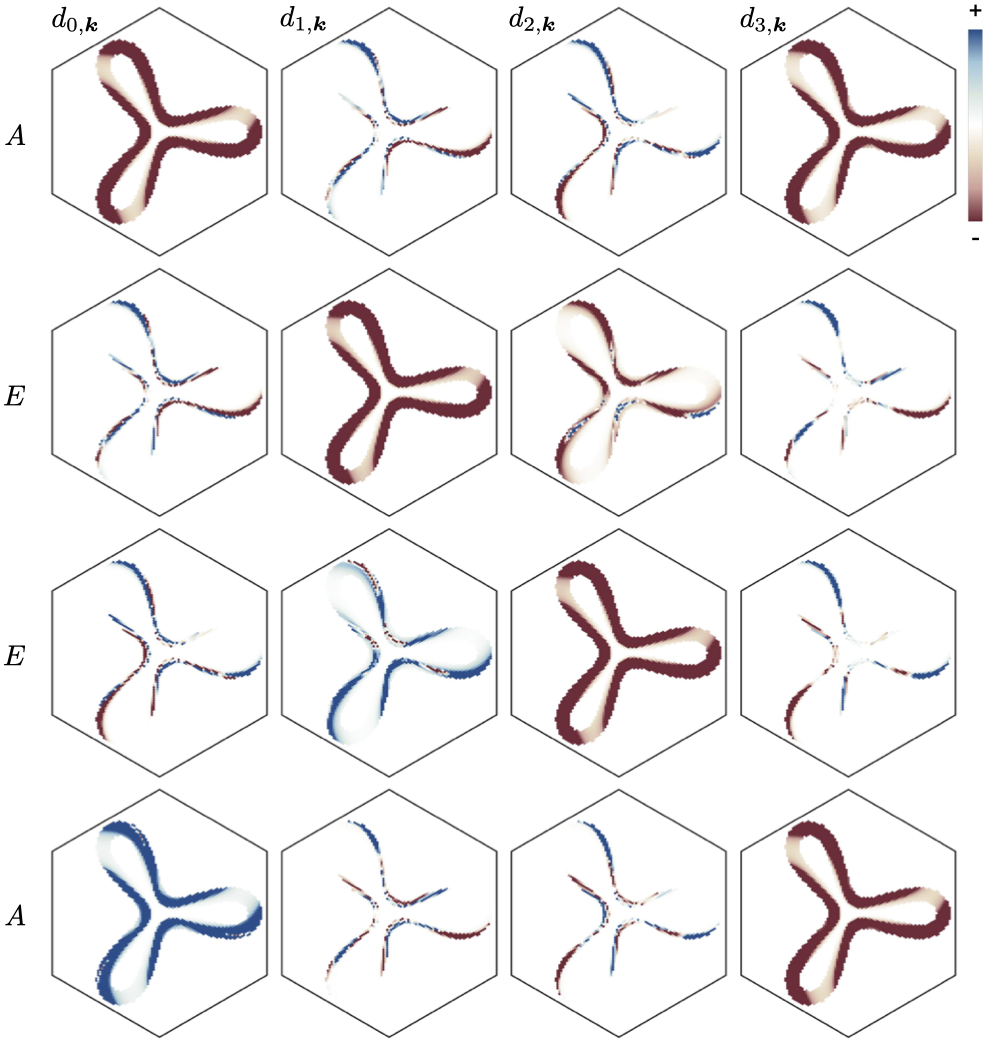}
    \caption{Spin-triplet favored interaction $\delta\gamma/\gamma_0=-0.25$, with fixed $D_0=6$ meV, $\lambda_{\text{R}}=\lambda_{\text{I}}=30$ meV  and $\varepsilon_F=9.6$ meV. (Top-to-bottom) Corresponds to leading eigenvalue down to fourth eigenvalue; the corresponding IRs are labeled.}
    \label{f:vecsSO3_RI}
\end{figure}

Using the linearized gap equation \eqref{gapeq} with vertex $\Gamma^{\text{even}}$ given by the first line of \equref{Gamma_eqns}, we compute the evolution of the superconducting order parameters, $d_{\mu, \vec k} = (d_{\vec{k},\eta=+})_\mu$, under applied SOC and displacement field, as well as for the different spin-symmetries of the parent superconducting state. The leading eigenvalues are presented in \figref{f:eigs} and selected eigenvectors can be found in Figs.~\ref{f:vecsSO3_R} and \ref{f:vecsSO3_RI}. In these plots, we have set $\theta=1.50^\circ$ and ${w_0/w_1=0.875}$ (which are tunneling parameters defined in \appref{A:Bandstr}), while for the gap equation \eqref{gapeq}, we keep eight bands $n$ (of a given valley) in the summation in ${\cal W}_{\mu\nu,\vec p}$. We now comment on the key features of these results:

\begin{figure}[t]
   \centering
    \includegraphics[width=\linewidth]
    {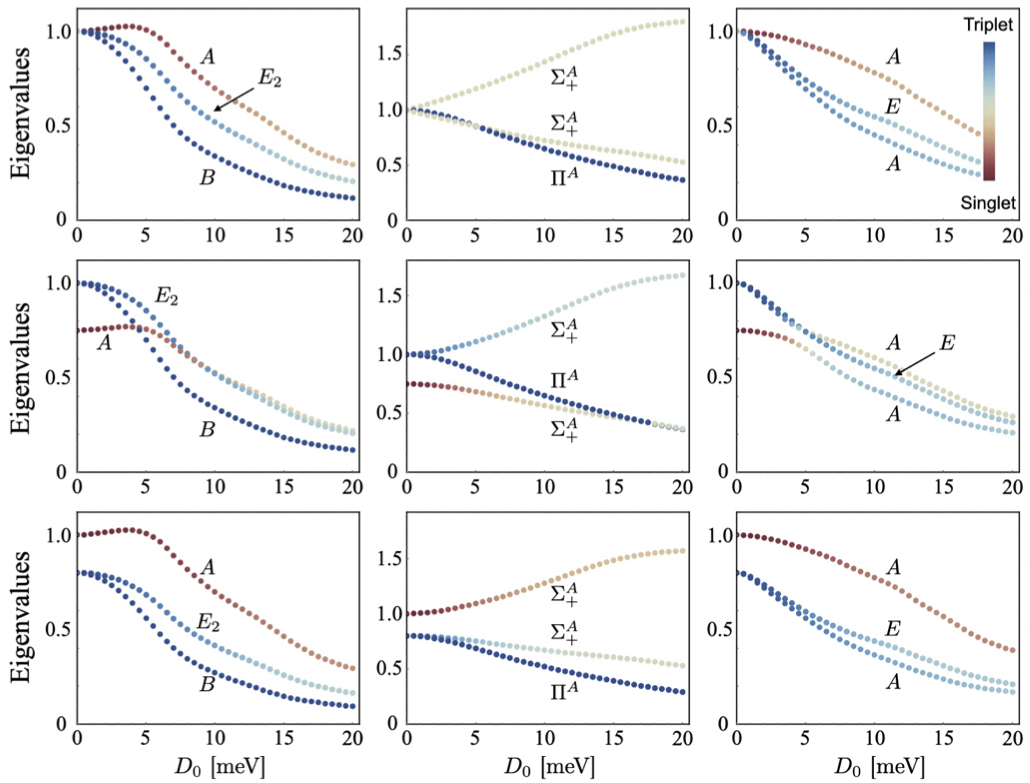}
    \caption{Eigenvalues [normalized to unity at $D_0=0$] as a function of $D_0$, (left column) $\lambda_{\text{R}}=30$ meV; (middle column) $\lambda_{\text{I}}=30$ meV; and (right column) $\lambda_{\text{I}}=\lambda_{\text{R}}=30$ meV. The distinct IRs are indicated, see \figref{fig:EvolutionofPairingAndSyms}(b) for left and right column. When $\lambda_{\text{R}}=0$, $\lambda_{\text{I}}\neq 0$ (middle column), the point group is $C_3 \times D_\infty$, see \figref{fig:EvolutionofPairingAndSyms}(a); we denote the two-dimensional (one-dimensional) IR which is trivial in $C_3$ and transforms as $x,y$ (also trivial) under $D_{\infty}$ by $\Pi^A$ ($\Sigma_+^A$). 
 (Top row) SO(4) symmetric parent state, i.e., $\delta\gamma=0$ in \equref{Gamma_eqns}.  
 (Middle row) Spin-triplet favored parent state, with interaction $\delta\gamma=-0.25$.  (Bottom row) Spin-singlet favored parent state, with interaction $\delta\gamma=0.25$. Everywhere the Fermi energy is $\varepsilon_F=9.6$ meV.}
    \label{f:eigs}
\end{figure}

\begin{enumerate}
\item Left column Fig.~\ref{f:eigs}: At $\lambda_{\text{R}}\neq0$ increasing $D_0$ generates a splitting of the states into the distinct IRs $A$, $B$, $E_2$ of $C_6^s$ [cf.~\figref{fig:EvolutionofPairingAndSyms}(b)].  We see from Fig.~\ref{f:vecsSO3_R} that the singlet $A$ has admixed in-plane triplet character, while the in-plane triplet $E_2$ has admixed singlet character; the admixture increases with $D_0$. The admixed components exhibit sign changes around the Fermi surface. The triplet $B$ (non-degenerate eigenvector),
however, remains a pure triplet with triplet vector
pinned to the out-of-plane direction. These features are in perfect agreement with the symmetry analysis, see Eqs.~(\ref{BuTriplet}), (\ref{SingletWithTriplet}) and (\ref{E2State}), and are further elucidated in \secref{SC_minimal} using a complementary description of the gap equation. That analysis will also explain why the admixed singlet (triplet) component(s) of the $E_2$ ($A$) state change sign between the two Fermi surfaces.

\item Middle row, left column Fig.~\ref{f:eigs}: For the case of the triplet preferred parent state $\delta\gamma<0$, and with $\lambda_{\text{R}}\neq0$, we  see that the leading eigenvalue exhibits a change from $E_2$ to $A$ at $D_0\approx9$ meV for $\lambda_{\text{R}}=30\,\textrm{meV}$. Since all three states transform according to different IRs, their eigenvalues cross each other (rather than exhibit avoided crossings), which corresponds to a phase transition. However, the precise form of the phase diagram---with or without phase coexistence---goes beyond the linearized gap equation and will be discussed in \secref{s:PT} below. 

\item Middle Column Fig.~\ref{f:eigs}: At $\lambda_{\text{I}}\neq0$ increasing $D_0$ generates a splitting of the states into the distinct IRs $\Sigma_+^A$,  $\Pi^A$ of $C_3 \times D_\infty$, with two non-degenerate states belonging to $\Sigma_+^A$, which correspond to the symmetric and antisymmetric combinations, i.e. $d_{0,\vec k}\pm d_{3,\vec k}$.

\item Right Column Fig.~\ref{f:eigs}: At $\lambda_{\text{R}},\lambda_{\text{I}}\neq0$ increasing $D_0$ generates a splitting of the states into the distinct IRs $A$, $E$ of $C_3^s$, again with two non-degenerate states belonging to $A$. These two non-degenerate states can be thought of as the ``bonding'' and ``anti-bonding'' configurations of the $A$ and $B$ states of $C_6^s$ in \figref{fig:EvolutionofPairingAndSyms}(b). This is confirmed by noting in the first (last) row in \figref{f:vecsSO3_RI} that the two singlet states $A$ have symmetric (antisymmetric) combinations of $d_{0,\vec k}$ and $d_{0,\vec k}$, i.e. different signs of $\beta$ in \equref{SingletRI}. This hybridization is also visible in the $D_0$ dependence of the eigenvalues in the last column of \figref{f:eigs}, which exhibits an avoided crossing between the associated pair of eigenvalues. The $A$ states also have admixed in-plane triplet character, with sign changes around the Fermi surface. On the other hand, the in-plane triplet $E$ (the second and third rows of \figref{f:vecsSO3_RI}) has admixed $d_{0,\vec k}$ and $d_{3,\vec k}$, which exhibit sign changes on the Fermi sheets. These features are well explained by the symmetry arguments [cf.~\equsref{SingletRI}{TripletRI}].

\end{enumerate}

\subsection{Phase transition}\label{s:PT}
\begin{figure}[tb]
\includegraphics[width=0.48\textwidth,clip]
{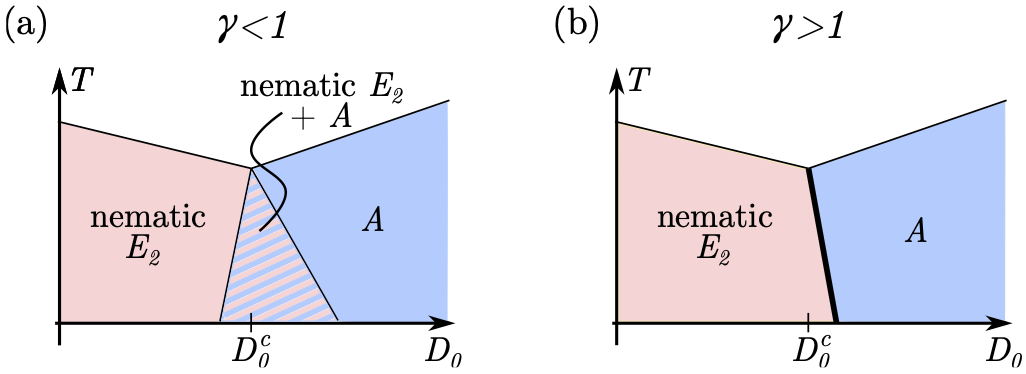}
 \caption{For $b_{E_2,2} < 0$, we either get (a) an intermediate region of microscopic coexistence or (b) a first order transition between the two superconductors, depending on the value of $\gamma$ in \equref{DefinitionOfGamma}.
 }
\label{fig:CoexistPhaseDiagrams}
\end{figure}

As already mentioned above, the crossing at a critical value $D_0=D_0^c$ between the leading and subleading eigenvalues in the middle row, left column panel in \figref{f:eigs} implies that there is a transition from the ($E_2$ of $C_{6}^s$) in-plane triplet with admixed singlet component in \equref{E2State} to the singlet with admixed in-plane triplet components of \equref{SingletWithTriplet}, transforming as $A$ of $C_{6}^s$. The linearized gap equation, however, does not yet fully determine the phase diagram in the vicinity of $D_0^c$ and $T_c$. To discuss this further, we expand the superconducting order parameter in \equref{SuperconductingOrderParameter} as
\begin{equation}
    d_{\vec{k},\eta} = \Psi \, d^A_{\vec{k},\eta} + \sum_{j=1,2}\phi_j\, d^j_{\vec{k},\eta}, \quad \Psi, \phi_j \in \mathbb{C}, \label{ExpansionInPsiAndPhi}
\end{equation}
where $d^A_{\vec{k},\eta}$ and $d^j_{\vec{k},\eta}$ are given by the four-component vectors in \equsref{SingletWithTriplet}{E2State}, respectively. Considering terms up to quartic order in $\Psi$ and $\vec{\phi}=(\phi_1,\phi_2)$ that are consistent with the $C_6^s$ point group, time-reversal, and gauge invariance, the free-energy must have the form
\begin{align}\begin{split}
    \mathcal{F} &\sim \frac{1}{2} a_A |\Psi|^2 + \frac{1}{2} a_{E_2} \vec{\phi}^\dagger\vec{\phi} + \frac{1}{4} b_A |\Psi|^4 + \frac{1}{4} b_{E_2,1}  (\vec{\phi}^\dagger \vec{\phi})^2 \\ & + \frac{1}{4} b_{E_2,2}  |\vec{\phi}^T \vec{\phi}|^2 + \frac{1}{2} c_1\, |\Psi|^2 \vec{\phi}^\dagger\vec{\phi} + \frac{1}{2} c_2 \,\text{Re} [\Psi^2 \vec{\phi}^\dagger\vec{\phi}^*] \label{FreeEnergyExpansion}
\end{split}\end{align}
close to $T_c$. Before proceeding, we first obtain an estimate for the real-valued coefficients of the quartic terms, $b_A$, $b_{E_2,1}$, $b_{E_2,2}$, $c_1$, $c_2$; we take the effective two-band normal-state Hamiltonian in \equref{FirstEffectiveHamiltonian}, neglect the spin splitting, $\vec{g}_{\vec{k}}=0$, and integrate out the fermions coupled to the superconducting order parameters $\Psi$, $\vec{\phi}$ via \equsref{ExpansionInPsiAndPhi}{SuperconductingOrderParameter}. Further neglecting the small admixture, $\alpha_{1,2} = 0$, and taking $\lambda, M = \text{const.}$ in \equsref{SingletWithTriplet}{E2State}, we find
\begin{equation}
    b_A = b_{E_2,1}/2 = -b_{E_2,2} = c_1/2 = c_2 >0, \label{EstimateOfTheCoefficients}
\end{equation}
where $b_A = 64 \sum_{\omega_n} \int\frac{\diff^2\vec{k}}{(2\pi)^2} (\omega_n^2 + \xi_{\vec{k}}^2)^{-2}$. Most importantly, we see that $b_{E_2,2} < 0$, favoring the unitary (nematic $E_2$) state, $\vec{\phi} \propto \hat{\vec{e}}_1$, over the non-unitary (chiral $E_2$) configuration, $\vec{\phi} \propto \hat{\vec{e}}_1 + i \hat{\vec{e}}_2$, where $\hat{\vec{e}}_{1,2}$ are two orthogonal $2$-component unit vectors. Minimizing the free energy in \equref{FreeEnergyExpansion} for $b_{E_2,2} < 0$, one finds that depending on 
\begin{equation}
    \gamma = \frac{c_1 - |c_2|}{\sqrt{b_A(b_{E_2,1}+b_{E_2,2})}}, \label{DefinitionOfGamma}
\end{equation}
one either obtains microscopic coexistence of $\Psi$ and nematic $\vec{\phi}$ ($\gamma < 1$) or a first-order transition between the two ($\gamma > 1$), see \figref{fig:CoexistPhaseDiagrams}. Interestingly, for the approximate estimate in \equref{EstimateOfTheCoefficients} we get $\gamma=1$ and these two possibilities are degenerate. As such, additional corrections coming from non-trivial form factors ($\lambda, M \neq \text{const.}$), SOC, and fluctuation- \cite{PhysRevLett.111.127001,PhysRevB.99.144507,OurClassification} or disorder \cite{Hoyer2014} corrections can determine which of the two scenarios in \figref{fig:CoexistPhaseDiagrams} is realized. In the case of microscopic coexistence, $\gamma < 1$, the relative phase of $\Psi$ and $\phi_j$ is determined by the sign of $c_2$. For the positive sign of $c_2$ obtained from our estimate in \equref{EstimateOfTheCoefficients}, we find $\vec{\phi}^* \Psi$ to be purely imaginary, such that the superconducting state in the hatched region in \figref{fig:CoexistPhaseDiagrams}(a) breaks time-reversal symmetry.

We note that significant corrections to \equref{EstimateOfTheCoefficients}, e.g., resulting from strong ferromagnetic fluctuations \cite{OurClassification}, can change the sign of $b_{E_2,2}$. Positive $b_{E_2,2}$ will favor $\vec{\phi} \propto \hat{\vec{e}}_1 + i \hat{\vec{e}}_2$ for sufficiently small $D_0$. However, close to $D_0 = D_0^c$, the competition between the terms $\propto b_{E_2,2}$ and $\propto c_2$ in \equref{FreeEnergyExpansion} will lead to more complex behavior than in \figref{fig:CoexistPhaseDiagrams}: depending on parameters, we find (not shown) either no coexistence, or coexistence with a first order transition, or a coexistence region with two different regimes (coexistence with only nematic or nematic and chiral pairing).

Since the transition involves superconducting states with different symmetries, there should be multiple ways of probing it experimentally.
Given the current experimental status in graphene moir\'e systems, we believe that the most straightforward way proceeds by measuring the variation of the superconducting critical temperature $T_c$ as a function of magnetic field $\vec{B}$ and displacement field $D_0$: the order parameter of the $A$ state does not allow to construct a time-reversal-odd, gauge-invariant composite order parameter and, hence, cannot couple linearly to the magnetic field $\vec{B}$ (even in the presence of strain). This is different for the $E_2$ order parameter since we can define $i( \phi_1^*\phi_2^{\phantom{*}} - \phi_2^*\phi_1^{\phantom{*}} )$, which transforms in the same way as the magnetic field component along the out-of-plane $z$ direction; it can, hence, couple linearly to it. In the presence of strain, it can also couple linearly to the in-plane magnetic field components. Consequently, the behavior of $T_c(\vec{B})$ is very different in the two phases and we therefore expect a drastic change of it when $D_0$ changes across the transition, making it directly observable.

\subsection{Character of the admixed state}\label{SC_minimal}
To further supplement the numerical studies of the gap equation \eqref{gapeq}, here we provide a complementary analysis to elucidate key features of the superconducting order parameters $(d_{\vec k,\eta})_\mu$. We specialize to $\lambda_\text{R}, D_0\neq0$ for this analysis.

We describe the behavior of the eigenvectors of the gap equation \eqref{gapeq} in terms of the effective $\vec{g}_{\eta\vec{k}}$-vector in \equref{FirstEffectiveHamiltonian}, which captures the impact of SOC (and $D_0$) on the partially-filled moir\'e bands at the Fermi level. The $\vec{g}_{\eta\vec{k}}$-vector can be explicitly computed from the full continuum model and its form can be inferred from \figref{fig:BS_Spin}. Recasting the mean-field Hamiltonian \eqref{supp_HBdG} in terms of $\vec{g}_{\eta\vec{k}}$ and separating ${\cal H}_\eta={\cal H}_\eta^{(0)} + {\cal H}_\eta^{(1)}$, 
\begin{align}
\label{Heff}
    &{\cal H}_\eta^{(0)}=\sum_{\vec k,s,s'} \left[\xi_{\eta\vec k}s_0 + \eta \vec g_{\eta\vec k}\cdot\vec s\right]_{s,s'}  c^\dag_{\vec k,s,\eta} c^\pdagger_{\vec k,s',\eta}\\
       &{\cal H}_\eta^{(1)}= \sum_{\vec k_1, \vec k_2;\mu,\nu} (-\Gamma^{-1})_{\vec k_1, \vec k_2; \mu,\nu} \bigl(d_{\vec k_1,\eta}^*\bigr)_\mu \bigl(d_{\vec k_2,\eta}\bigr)_\nu\, \\
     \notag &+ \sum_{\vec k;s_1,s_2;\mu}\Big\{c_{ \eta,n,\vec k}^\dag c_{\eta,n',-\vec k}^\dag\left[(d_{\vec k,\eta})_\mu (s_\mu i s_y )_{s_1,s_2}\right]   +  \text{H.c.}\Big\}.
\end{align}
The eigenvalues of ${\cal H}_\eta^{(0)}$  are $E_{\pm,\eta\vec k}=\xi_{\eta\vec k} \pm |\vec g_{\eta\vec k}|$ and the subsequent normal-state Greens function is conveniently expressed as
\begin{align}
\label{Geff}
    {\cal G}^{\eta,\eta'}_{\vec k, i\omega_n}&=\left[G^+_{\eta\vec k, i\omega_n} s_0 + \eta \hat{\vec g}_{\eta\vec k}\cdot \vec s \ G^-_{\eta\vec k, i\omega_n} \right] \delta_{\eta,\eta'},
\end{align}
with $ \hat{\vec g}_{\eta\vec k}= \vec g_{\eta\vec k}/| \vec g_{\eta\vec k}|$ and
\begin{align}
    G^\pm_{\eta\vec k, i\omega_n}&=\left[ \frac{1}{i\omega_n - E_{+,\eta\vec k}} \pm \frac{1}{i\omega_n - E_{-,\eta\vec k}}  \right].
\end{align}

We consider an interaction vertex in the superconducting channel, $\Gamma_{\vec k,\vec k';\mu\nu}=\Gamma^0_{\vec k,\vec k'}\delta_{\mu\nu}$, which strictly favors intervalley pairing, and which is even in the quasi-momentum indices, i.e.~$\Gamma^0_{\vec k,\vec k'}=\Gamma^0_{-\vec k,\vec k'}=\Gamma^0_{\vec k,-\vec k'}=\Gamma^0_{-\vec k,-\vec k'}$; to be explicit we take the interaction as given by the first line in \equref{Gamma_eqns} in the limit $\delta \gamma = 0$. Utilizing that the Greens function is diagonal in valley indices, as per \equref{Geff}, the linearized gap equation reads
\begin{align}
\label{Gap_Eqn_Minimal_Full}
    \notag &(\Delta_{\vec k}^{\eta,-\eta})_{s,s'}=T\sum_{n,\vec k'; s_1,s_2;\eta}\Gamma^0_{\vec k,\vec k'} \times \\
    & ({\cal G}^{\eta,\eta}_{\vec k', i\omega_n})_{s,s_1}(\Delta^{\eta,-\eta}_{\vec k'})_{s_1,s_2} ({\cal G}^{-\eta,-\eta}_{-\vec k', -i\omega_n})_{s',s_2}.
\end{align}
As before in \secref{NumericalResultsGapped}, we specialize to a given $\eta=+$ and denote $d_{\mu, \vec k}=(d_{\vec k, +})_\mu$ such that the decomposition into the spin-singlet and triplet components is,
\begin{align}
    \Delta^{+-}_{\vec k}&=d_{0,\vec k}s_0 + \vec d_{\vec k}\cdot \vec s.
\end{align}
The gap equation \eqref{Gap_Eqn_Minimal_Full} reduces to 
\begin{align}
\notag &d_{0,\vec k}=\sum_{\vec k'} \Gamma_{\bm k,\bm k'}^0\left[{\mathbb V}^+_{T,\vec k'}d_{0,\vec k'} + {\mathbb V}^-_{T,\vec k'}\hat{\vec g}_{\vec k'}\cdot \vec d_{\vec k'}\right],\\
\label{triplet_gapeqn}
 &\vec d_{\vec k}=\sum_{\vec k'} \Gamma_{\bm k,\bm k'}^0\bigg[{\mathbb V}^+_{T,\vec k'}\vec d_{\vec k'}+ {\mathbb V}^-_{T,\vec k'}\hat{\vec{g}}_{\vec k'}d_{0,\vec k'} \\
\notag &\hspace{1cm} +2 {\mathbb V}^0_{T,\vec k'}[\hat{\vec g}_{\vec k'}(\hat{\vec g}_{\vec k'}\cdot\vec d_{\vec k'})-\vec d_{\vec k'}] \bigg].
\end{align}
We have defined the sums and differences of thermal occupation factors,
\begin{align}
\label{fpm}
{\mathbb V}^\pm_{T,\vec k}&=\frac{1}{2}T\sum_n \left[ \frac{1}{E_{+,\vec k}^2+\omega_n^2} \pm  \frac{1}{E_{-,\vec k}^2+\omega_n^2} \right],\\
{\mathbb V}^0_{T,\vec k}&=
 T \sum_n \frac{(E_{+,\vec k}-E_{-,\vec k})^2}{(E_{+,\vec k}^2+\omega_n^2)(E_{-,\vec k}^2+\omega_n^2)}.
\end{align}

To understand the important characteristics of the superconducting order parameters $d_{\mu, \mathbf{k}}$, which we observed in \secref{NumericalResultsGapped} through a comprehensive numerical analysis of the gap equation \eqref{gapeq}, we will now present our analytic findings based on a perturbative expansion in ${\mathbb V}^-_{T,\mathbf{k}}$. Specifically, we will examine the properties of the $A$ and $E_2$ states and compare our results with the findings presented in \figref{f:vecsSO3_R}. The explicit steps of the perturbative expansion are provided in \appref{PerturbationTheory}.

It is important, for what follows, to first establish the (approximate) equality 
\begin{equation}
   {\mathbb V}^\pm_{T,\vec k_{F}^+}=\pm {\mathbb V}^\pm_{T,\vec k_{F}^-}, \label{VIdentity}
\end{equation}
where we define $\vec k_{F}^\pm$ as points on the Fermi surfaces for the $E_{\pm,\vec k}$ bands, i.e.~$E_{\pm,\vec k_{F}^\pm}=0$. To this end, we assume $|\vec g_{\vec k}|$ does not change between Fermi surfaces, i.e. $|\vec g_{\vec k_{F}^+}|=|\vec g_{\vec k_{F}^-}|\equiv|g|$. Next, employ a linear expansion of the dispersion in momentum normal to a point $\vec{k}_{F}^0$  on the Fermi surface of $\xi_{\vec{k}}$, i.e. denoting the normal vector $\hat{n}^{FS}_{\vec{k}_{F}^0}$, we expand in $\delta\vec{k}= (k - k_{F}^0)\hat{n}^{FS}_{\vec{k}_{F}^0}$ such that $\xi_{\delta\vec{k}}\approx v (k - k_{F}^0)$. From $E_{\pm,\vec k_{F}^\pm}=0$ we see that $k_{F}^\pm=k_{F}^0 \pm |g|/v$ and so $E_{+,\vec k_{F}^-}^2=E_{-,\vec k_{F}^+}^2=4|g|^2$; this implies ${\mathbb V}^\pm_{T,\vec k_{F}^+}=\pm {\mathbb V}^\pm_{T,\vec k_{F}^-}$.

\vspace{1em}\noindent{\bf $A$ state.}
We start by considering the dominant singlet state $d_{0,\vec k}$, with admixed in-plane triplet components $\vec d^\parallel_{\vec k}=(d_{1,\vec k}, d_{3,\vec k})$-vector (and ignoring the $d_{3,\vec k}$ component which is decoupled). Reinstating the valley index $\eta$, we denote the unperturbed (i.e. ${\mathbb V}^-_{T,\bm k}\to0$) $A$-state as 
\begin{align}
d^A_{\eta\cdot\bm k} & = (d^{(0)}_{0,\eta\cdot\bm k}; \bm 0),
\end{align}
with $d^{(0)}_{0,\eta\cdot\bm k}$ approximately constant in $\bm k$. The perturbation ${\mathbb V}^-_{T,\eta\cdot\bm k}$ generates admixed triplet components,
\begin{align}
d^A_{\eta\cdot\bm k}{}'& =  (d^{(0)}_{0,\eta\cdot\bm k}; \beta \eta {\mathbb V}^-_{T,\eta\cdot\vec k} \ \hat{\vec{g}}_{\eta\cdot\vec k}),
\end{align}
with $\beta$ a dimensionless factor (see \appref{PerturbationTheory} for details). Since the two components of $\hat{\vec g}_{\eta\cdot\vec k}$ transform as $k_x$ and $k_y$ under $C_{3z}$, this is consistent with the form in \equref{SingletWithTriplet} predicted by symmetry.  Moreover, this simple expression already captures the key features clearly seen in the full numerical results in the third line of \figref{f:vecsSO3_R}; the admixed triplet vector changes sign between the spin-split Fermi surfaces due to the factor ${\mathbb V}^-_{T,\eta\cdot\vec k}$ and \equref{VIdentity}. We further see that it inherits the directional dependence of $\hat{\vec g}_{\eta\cdot\vec k}$, including its M\"obius texture \cite{MobiusPRL} for $\lambda_{\text{I}} \ll \lambda_{\text{R}}$, see \figref{fig:BS_Spin}(b).

\begin{figure*}[t]
   \centering
    \includegraphics[width=0.8\linewidth]
    {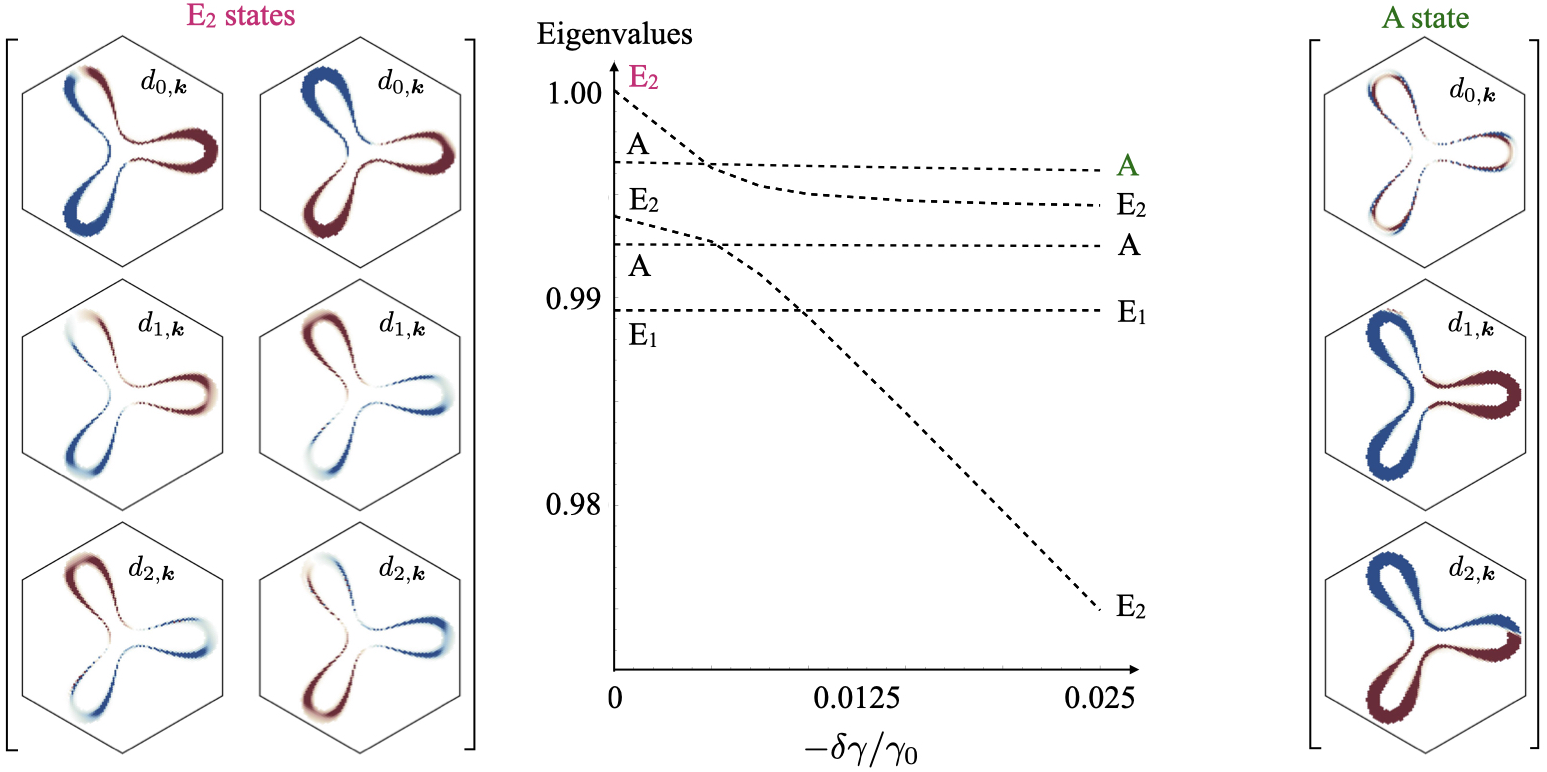}
    \caption{Nodal pairing: states and their evolution with $\delta\gamma$. Dashed lines are the eigenvalues obtained by direct computation of the gap equation \eqref{gapeq}. Eigenvectors for the leading $E_2$ state at $\delta\gamma/\gamma_0=0$ are shown on the left hand side, while those for the leading $A$ state at $\delta\gamma/\gamma_0=-0.025$ are on the right hand side. Parameters $\{\lambda_R, D_0, \varepsilon_F\}=\{6, 6, 9.6\}$ meV.}
    \label{f:NodalPairing}
\end{figure*}

\vspace{1em}\noindent{\bf $E_2$ states.}
Similar to above, we denote the unperturbed (i.e. ${\mathbb V}^-_{T,\eta\cdot\bm k}\to0$) $E_2$-state as
\begin{align}
d^{E_2}_{\eta, \bm k} & = (0; \eta \bm d^{(0)}_{\eta\cdot \bm k}),
\end{align}
where, to a good approximation, $\bm d^{(0)}_{\eta\cdot\bm k} \approx {\bm e}_\parallel$, with ${\bm e}_\parallel$ a constant in-plane vector, for all ${\bm k}$ such that $\Theta_{\bm k; \varepsilon_F}=1$, i.e. in the grid of Fig. \ref{fig:FSgrid}. The perturbation ${\mathbb V}^-_{T,\eta\cdot\bm k}$ generates admixed singlet components,
\begin{align}
d^{E_2}_{\eta,\bm k}{}' & =  (\beta  {\mathbb V}^-_{T,\eta\cdot\vec k} \eta \hat{\vec{g}}_{\eta\cdot\vec k}\cdot {\bm e}_\parallel;  \ \eta\bm d^{(0)}_{\eta\cdot\bm k}),
\end{align}
This first-order perturbation expansion already captures the key features observed in the $E_2$ states presented in \figref{f:vecsSO3_R}; the admixed singlet component changes sign between the spin-split Fermi surfaces due to the factor ${\mathbb V}^-_{T,\eta\cdot\vec k}$ obeying \equref{VIdentity} and further inherits the directional dependence of $\hat{\vec g}_{\eta\cdot\vec k}\cdot {\bm e}_\parallel$.

\subsection{Nodal pairing}\label{s:SC_nodal}
In this final section, we start from a nodal parent state and examine its characteristics under applied $D_0$ and SOC. To this end, \figref{f:NodalPairing} illustrates the evolution of the distinct IRs as the singlet-triplet interaction-asymmetry parameter $\delta \gamma$ in \equref{Gamma_eqns} is varied. The numerical results of \figref{f:NodalPairing} are obtained from the gap equation \eqref{gapeq} by using the odd-vertex $\Gamma^\text{odd}_{\vec k_1, \vec k_2; \mu,\nu}$ of \equref{Gamma_eqns}, which enables nodal pairing.
As can be seen from the eigenvalues, the system is in an $E_2$ state at $\delta\gamma/\gamma_0=0$, and subsequently undergoes a crossing with an $A$ state (signaling a phase transition, c.f.~Sec.~\ref{s:PT}) as $-\delta\gamma/\gamma_0$ is increased. The dominant eigenvector(s) at each limit of  $\delta\gamma/\gamma_0$ are also shown---and demonstrate a switch in behavior from a singlet-dominant $E_2$ state to that of a triplet-dominant $A$ state.  
To complete our analysis of the nodal pairing, we detail the structure of each of the distinct states $\{A, A, E_2, E_2, E_1\}$ of \figref{f:NodalPairing}, focusing on the region $\delta\gamma/\gamma_0\approx-0.025$. We show how the order parameter structures relate to the known vectors of the system, i.e.~the $\vec{g}_{\eta\vec{k}}$ coming from SOC; the $({\cal X}_{\eta\vec{k}}, {\cal Y}_{\eta\vec{k}})$ coming from the interaction vertex \equref{Gamma_eqns}; and the vector triad $\{\hat{\vec e}_x,\hat{\vec e}_y, \hat{\vec e}_z\}$ for the spin-triplet components. 

\vspace{1em}\noindent{\bf $A$ states.} Considering the first (highest eigenvalue) $A$ state of \figref{f:NodalPairing} at $\delta\gamma/\gamma_0\approx-0.025$, the numerical solution reveals the following structure
\begin{equation}
    d_{\vec{k},\eta} = \beta \ (\lambda_{\eta\vec{k}};0,0,0) +   \eta (0;{\cal X}_{\eta\vec{k}},{\cal Y}_{\eta\vec{k}},0). \label{A1}
\end{equation}
Here $\lambda_{\eta\vec{k}}$ is not close to the identity, but is instead approximately of the form $\lambda_{\eta\vec{k}}\approx\vec{g}_{\eta\vec{k}}\cdot ({\cal X}_{\eta\vec{k}}, {\cal Y}_{\eta\vec{k}}) {\mathbb V}^-_{T,\eta\vec k}$, which transforms like a scalar under three-fold rotations. As a reminder, the function ${\mathbb V}^-_{T,\eta\vec k}$ changes sign between the two Fermi surfaces, see \equref{VIdentity}, and $\beta$ is a small dimensionless parameter. Meanwhile, the second highest $A$ state of \figref{f:NodalPairing} takes a rather different form
\begin{equation}
    d_{\vec{k},\eta} =    \beta \ (\tilde{\lambda}_{\eta\vec{k}};0,0,0)  +    \eta (0;{\cal Y}_{\eta\vec{k}},-{\cal X}_{\eta\vec{k}},0). \label{A2}
\end{equation}
Here $\tilde{\lambda}_{\eta\vec{k}}\approx \vec{g}_{\eta\vec{k}}\cdot ({\cal Y}_{\eta\vec{k}}, -{\cal X}_{\eta\vec{k}}){\mathbb V}^-_{T,\eta\vec k}$ is again a scalar under three-fold rotations and the triplet components of \eqref{A2} are seen to be simply a $\pi/2$-rotation relative to those of \eqref{A1}. These two $A$ states can be understood as resulting from the rotationally trivial combination of $({\cal X}_{\eta\vec{k}}, {\cal Y}_{\eta\vec{k}})$ with each of the two components of the nodeless $E_2$ state of \figref{f:vecsSO3_R}. Within our symmetry analysis in \secref{SymmetryDiscussion2DIRs}, they correspond to the admixture of $A_g$ to $A_u$ in \equref{AuState}, see \figref{fig:EvolutionofPairingAndSyms}(b). This is consistent with triplet being dominant in both \equsref{A1}{A2}, while the singlet components are admixed. As per our discussion in \secref{SC_minimal} the admixed components exhibit sign changes between the Fermi surfaces.

\vspace{1em}\noindent{\bf $E_2$ states.} The highest $E_2$ states of \figref{f:NodalPairing} at $\delta\gamma/\gamma_0\approx-0.025$ exhibit the behavior
\begin{align}
  \notag \begin{pmatrix} d^1_{\vec{k},\eta} \\ d^2_{\vec{k},\eta}  \end{pmatrix} &= \beta_1 \ \begin{pmatrix} ({\cal X}_{\eta\vec{k}};0,0,0) \\ ({\cal Y}_{\eta\vec{k}};0,0,0) \end{pmatrix} + \beta_2 \ {\mathbb V}^-_{T,\eta\vec k} \begin{pmatrix} (g^x_{\eta\vec{k}};0,0,0) \\ (g^y_{\eta\vec{k}};0,0,0) \end{pmatrix} \\
  &+ \eta\begin{pmatrix} (0;{\cal X}_{\eta\vec{k}},-{\cal Y}_{\eta\vec{k}},0) \\ (0;-{\cal Y}_{\eta\vec{k}},-{\cal X}_{\eta\vec{k}},0) \end{pmatrix}, \label{E2_1}
\end{align}
with $\beta_1, \beta_2$ small dimensionless parameters; this expression implies that the singlet components of these $E_2$ states are a linear combination of $({\cal X}_{\eta\vec{k}}, {\cal Y}_{\eta\vec{k}})$ and $\vec g_{\eta\vec{k}}$; the explicit ratio depends on the strength of SOC.  
The in-plane triplet components of \eqref{E2_1} are dominant, while the singlet components are admixed. The next set of $E_2$ states in \figref{f:NodalPairing} take the form, e.g., 
\begin{align}
 \notag \begin{pmatrix} d^1_{\vec{k},\eta} \\ d^2_{\vec{k},\eta}  \end{pmatrix} &\approx \begin{pmatrix} ({\cal X}_{\eta\vec{k}};0,0,0) \\ ({\cal Y}_{\eta\vec{k}};0,0,0) \end{pmatrix} + \beta_1 \ \eta\begin{pmatrix} (0;{\cal X}_{\eta\vec{k}},-{\cal Y}_{\eta\vec{k}},0) \\ (0;-{\cal Y}_{\eta\vec{k}},-{\cal X}_{\eta\vec{k}},0) \end{pmatrix}\\
   &\hspace{0.25em}+ \beta_2 \ \eta \ {\mathbb V}^-_{T,\eta\vec k} \begin{pmatrix} (0;g^x_{\eta\vec{k}} {\cal X}_{\eta\vec{k}},g^y_{\eta\vec{k}} {\cal X}_{\eta\vec{k}},0) \\ (0;g^x_{\eta\vec{k}} {\cal Y}_{\eta\vec{k}},g^y_{\eta\vec{k}} {\cal Y}_{\eta\vec{k}},0) \end{pmatrix}.
  \label{E2_2}
  \end{align}
In contrast to \eqref{E2_1}, the $E_2$ state in \equref{E2_2} can be understood as resulting from the rotationally non-trivial combination of the $({\cal X}_{\eta\vec{k}}, {\cal Y}_{\eta\vec{k}})$ with the single component the nodeless $A$ state of \figref{f:vecsSO3_R}. In this case the singlet components are dominant and the triplet components are admixed. With this insight, we comment that the triplet components of \eqref{E2_2} with coefficient $\beta_2$ are deduced as being the rotationally-trivial combination $[(\hat{\vec e}_x,\hat{\vec{e}}_y,0)\cdot(\vec g^x_{\eta\vec{k}}, \vec g^y_{\eta\vec{k}},0)]$---as seen in the triplet components of the $A$ state of \figref{f:vecsSO3_R}---multiplied by the vector $({\cal X}_{\eta\vec{k}}, {\cal Y}_{\eta\vec{k}})^T$. In terms of our symmetry analysis in \secref{s:symm_analysis}, the states in \equsref{E2_1}{E2_2} should be thought of as two different superpositions of $E_{2g}$ in \equref{E2g1StateSynAn} and $E_{2u}$ in \equref{TripletPairingC6hs}, leading to \equref{GappingOutAdmixture} with non-trivial $M_{\vec{k}}$, see also \figref{fig:EvolutionofPairingAndSyms}(b).

\vspace{1em}\noindent{\bf $E_1$ states.}  Finally, there exists $E_1$ states, which are precisely those presented in \eqref{E1u}. 
These are purely out-of-plane triplets and do not mix with singlets or in-plane triplets if only Rashba SOC is present.

\section{Conclusion and Outlook}\label{s:Discussion}
We presented the continuum model for a TMD/TTLG/TMD heterostructure, subject to two distinct arrangements---(i) inversion symmetric and (ii) mirror symmetric, see top two panels in \figref{f:Bandstr}. This difference has significant implications for the form of the proximitised spin-orbit coupling (SOC); with (i) the SOC does not itself generate spin-splitting, but instead splitting can be induced via, e.g., an inversion-symmetry-breaking displacement field ($D_0$). In contrast, in (ii) an Ising SOC generates spin-splitting, already at $D_0=0$. These results are summarized in \figref{f:Bandstr}. The inversion symmetric setup, which hosts a displacement-field-tunable spin-splitting was the focus of the rest of the analysis. 

Our objective was to systematically understand how the superconductivity of TTLG, referred to as the parent superconducting state, evolves under applied $D_0$ and SOC, where we considered both Rashba and Ising SOC. We accounted for a range of possible candidate parent superconducting states, including spin-singlet, -triplet, and -SO(4) ordering, with either nodeless or nodal momentum dependence. We exclusively focused on intervalley pairing, as it is expected to be favored over intravalley pairing due to time-reversal symmetry and an additional protection against disorder \cite{OurMicroscopicTDBG}. To achieve this goal, we pursued a three-fold analysis. First, we applied a mean-field gap equation (detailed in \secref{s:SC_model}) to numerically compute the superconducting order parameters (see \secref{NumericalResultsGapped}), allowing for arbitrary momentum dependence. We supplemented this with analytic arguments to understand salient features (\secref{SC_minimal}), and we also provided a detailed symmetry analysis (\secref{s:symm_analysis}), which has the advantage of being independent of the particular assumptions---most notably the form of interactions---entering the gap equation and the mean-field approximation itself. These three approaches complement each other well and yield a consistent picture for the evolution of superconductivity in TTLG with $D_0$-tunable SOC. 

For dominant Rashba SOC (expected for $\theta_{\text{TMD}}$ in \figref{f:Bandstr} close to $30^\circ$), the system is predicted to exhibit phase transitions as a function of $D_0$ between two different pairing states if and only if triplet pairing is dominant in the parent TTLG system; this could be used to probe the spin structure of the superconducting state in TTLG in future experiments, e.g., by measuring the evolution of the critical temperature as a function of $D_0$ and magnetic field in our geometry (i). We have studied the phase diagram in the vicinity of the transition point, which either exhibits a first order transition or an intermediate regime of microscopic coexistence, see \figref{fig:CoexistPhaseDiagrams}.

Future theoretical studies may build upon the results obtained here to consider further implications of the pairing states. In particular, they may take the array of superconducting order parameters from this work and construct the gauge-invariant bilinears of the orders, known as {\it vestigial orders} \cite{VestigialOrderReview}. These orders break only a subset of the symmetries and often only discrete symmetries, allowing them to persist at temperatures above the critical phase coherence temperature of the underlying or constituent superconducting order. This persistence can influence the electronic behavior of the system outside of the superconducting phase. Given the exotic structure of the superconducting order parameters found here, one may hope to find exotic vestigial orders with clear experimental signatures.  Additionally, the topological nature of the superconducting order parameters has not been considered in this work and, therefore, remains an open problem. Related systems have demonstrated the existence of higher-order topological superconductivity \cite{LiInghamScammell2020, chew2021higherorder, LiHOTS, scammell2021intrinsic}, including for intervalley paired states \cite{LiHOTS, scammell2021intrinsic}. 

A direct extension of the present analysis would be to employ the inversion symmetric TMD/TTLG/TMD heterostructure, combined with inversion breaking $D_0$, to study the evolution of candidate particle-hole phases in this system. Also their interplay with superconductivity will likely give rise to very interesting and rich physics \cite{LiuReview2021,doi:10.1126/science.aaw3780,2021arXiv211207127S, Jiang-Xiazi_diode,2023arXiv230101344P,PhysRevB.106.235157,2023arXiv230317529C,2023arXiv230506949P, ingham2023quadratic}. Therefore, it would be instructive to analyze how they coexist or compete with superconductivity. 

The ultimate goal of this analysis is to provoke future experimental studies of this highly tunable van der Waals heterostructure, to systematically investigate the influence of spin-splitting of the electronic bands on correlated phases in general, and, in particular, to use our systematic results as a means of distinguishing between various candidate parent superconducting states.

\

\begin{acknowledgments}
M.S.S.~acknowledges funding from the European Union (ERC-2021-STG, Project 101040651---SuperCorr). Views and opinions expressed are however those of the authors only and do not necessarily reflect those of the European Union or the European Research Council. Neither the European Union nor the granting authority can be held responsible for them.

\end{acknowledgments}


\onecolumngrid

\begin{appendix}

\clearpage
\section{Band structure -- Extra details}\label{A:Bandstr}
\noindent
{\bf TTLG band Hamiltonian.}---As presented in Sec. \ref{ContinuumModel}, the continuum Hamiltonian is separated into four parts, $h_{\vec{k},\eta} = h^{(g)}_{\vec{k},\eta} + h^{(t)}_{\vec{k},\eta} + h^{(D)}_{\vec{k}} + h^{(\text{SOC})}_{\vec{k},\eta}$; the terms $h^{(D)}_{\vec{k}}$ and $h^{(\text{SOC})}_{\vec{k},\eta}$, which we think of as a perturbation in this work, were described in the main text, and here we provide the explicit form of the TTLG subsystem, i.e. the $h^{(g)}_{\vec{k},\eta} + h^{(t)}_{\vec{k},\eta}$. First, the contribution from the individual graphene layers is captured by,
\begin{align}    \left(h^{(g)}_{\vec{k},+}\right)_{\rho,\ell,s,\vec{G};\rho',\ell',s',\vec{G}'} &= \delta_{\ell,\ell'} \delta_{s,s'}\delta_{\vec{G},\vec{G}'} v_F (\vec{\rho}_{\theta_\ell})_{\rho,\rho'} \left(\vec{k} + \vec{G} - (-1)^\ell \vec{q}_{1}/2 \right), \label{DiracCones} \\ \left(h^{(g)}_{\vec{k},-}\right)_{\rho,\ell,s,\vec{G};\rho',\ell',s',\vec{G}'} &= \left(h^{(g)}_{-\vec{k},+}\right)^*_{\rho,\ell,s,-\vec{G};\rho',\ell',s',-\vec{G}'},
\end{align}
where $\vec{q}_1$ connects the $K$ and $K'$ points of the MBZ and $\vec{\rho}_{\theta} = e^{i \theta \rho_z/2} \vec{\rho} e^{-i \theta \rho_z/2}$. Second, the tunneling between the layers is modeled as,
\begin{align}\begin{split}
\left(h^{(t)}_{\vec{k},+}\right)_{\rho,\ell,s,\vec{G};\rho',\ell',s',\vec{G}'} &= \sqrt{2}\, \delta_{s,s'} \hspace{-0.2em} \begin{pmatrix} 0 & (T_{\vec{G}-\vec{G}'})_{\rho,\rho'} & 0 \\  (T_{\vec{G}'-\vec{G}}^*)_{\rho',\rho} & 0 & 0 \\ 0 & 0 & 0 \end{pmatrix}_{\ell,\ell'}, \hspace{-0.3em} \\ \left(h^{(t)}_{\vec{k},-}\right)_{\rho,\ell,s,\vec{G};\rho',\ell',s',\vec{G}'} &= \left(h^{(t)}_{-\vec{k},+}\right)^*_{\rho,\ell,s,-\vec{G};\rho',\ell',s',-\vec{G}'}, \label{TunnelingMatrixNotation}
\end{split}\end{align}
where,
\begin{align}
    T_{\delta\vec{G}} = \sum_{j=-1,0,1}\delta_{\delta\vec{G}+\vec{A}_j,0} \left[w_0 \rho_0 + w_1 \begin{pmatrix} 0 & \omega^j \label{FormOfT} \\  \omega^{-j} & 0 \end{pmatrix} \right], \\ \omega = e^{i \frac{2\pi}{3}}, \quad \vec{A}_0 =0, \quad \vec{A}_1 = \vec{G}_1, \quad \vec{A}_2 = \vec{G}_1 + \vec{G}_2. 
\end{align}
Here $w_0$ and $w_1$ parametrize the strength of, respectively, the sublattice diagonal and off-diagonal interlayer hopping strengths. All results presented here take $w_0 = 0.875w_1$, $w1 = 110$meV.

\begin{figure*}[t!]
   \centering
    \includegraphics[width=0.95\linewidth]
    {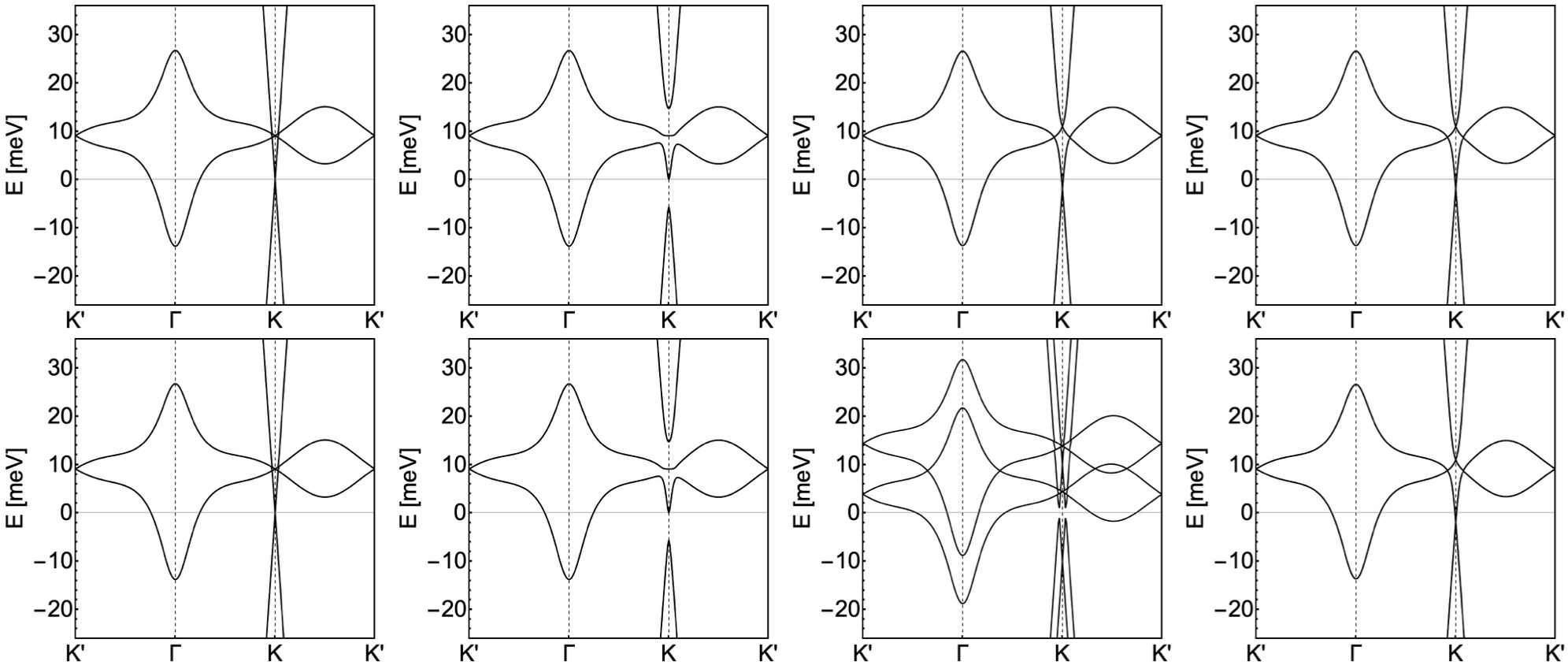}
    \caption{Band structure at twist $\theta=1.75^\circ$, with $\eta=1$. Top row: $\theta_\text{TMD}+\pi$ (inversion symmetric stack). Bottom row: $\theta_\text{TMD}$. The following combinations of SOC and $D_0$: (i) $\lambda_{\text{R}}, \lambda_{\text{I}}, D_0 = \{0,0,0\}$meV,  (ii) $\lambda_{\text{R}}, \lambda_{\text{I}}, D_0 = \{10,0,0\}$meV, (iii) $\lambda_{\text{R}}, \lambda_{\text{I}}, D_0 = \{0,10,0\}$meV, (iv) $\lambda_{\text{R}}, \lambda_{\text{I}}, D_0 = \{0,0,10\}$meV}
    \label{fig:BScompare1}
\end{figure*}

\begin{figure*}[h!]
   \centering
    \includegraphics[width=0.95\linewidth]
    {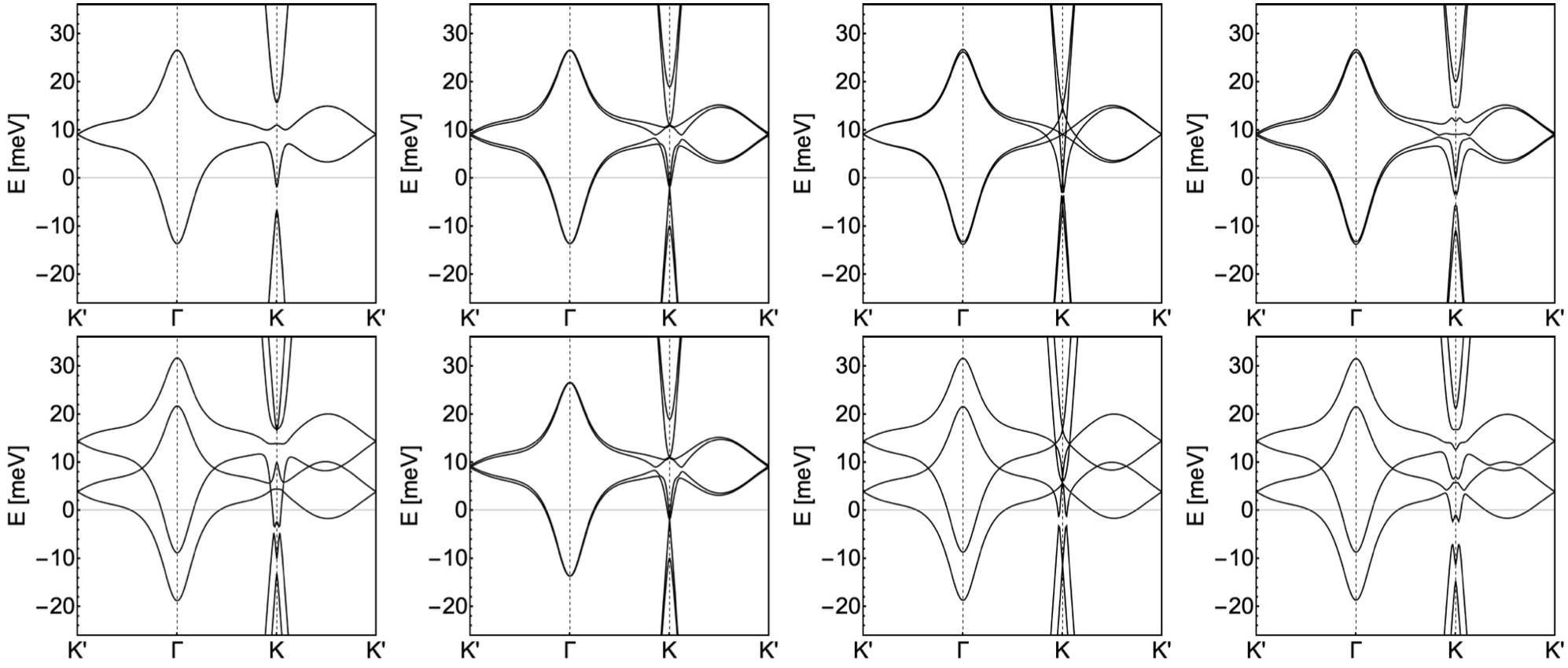}
    \caption{Band structure at twist $\theta=1.75^\circ$, with $\eta=1$. Top row: $\theta_\text{TMD}+\pi$ (inversion symmetric stack). Bottom row: $\theta_\text{TMD}$. The following combinations of SOC and $D_0$: (i) $\lambda_{\text{R}}, \lambda_{\text{I}}, D_0 = \{10,10,0\}$meV, (vi) $\lambda_{\text{R}}, \lambda_{\text{I}}, D_0 = \{10,0,10\}$meV, (vii) $\lambda_{\text{R}}, \lambda_{\text{I}}, D_0 = \{0,10,10\}$meV, (viii) $\lambda_{\text{R}}, \lambda_{\text{I}}, D_0 = \{10,10,10\}$meV}
    \label{fig:BScompare2}
\end{figure*}

\vspace{1em}\noindent
{\bf Extra plots.}---To supplement the band structure plots presented in \figref{f:Bandstr}, which compared the band structures of the two distinct configurations, here Figs.~\ref{fig:BScompare1} and \ref{fig:BScompare2} provide a more comprehensive comparison for a range of SOC and $D_0$ strengths. 

\clearpage

\section{Symmetrizing the Gap Equation}
\label{A_symmetrizing}
The gap equation \eqref{gapeq} is a non-Hermitian eigenvalue problem; the order parameters $d_{\mu, \bm k}$ presented in the main text are found  by solving for the right eigenvector of the non-Hermitian matrix. However, since it is convenient to work with Hermitian matrices -- especially for application of perturbation theory detailed in Appendix \ref{PerturbationTheory} -- we here show how to symmetrize the gap equation \eqref{gapeq}. Assuming summation over repeated indices, the gap equation is written as
  \begin{align}
 (\Gamma^{-1})_{\bm k_1,\bm k_2;\mu\nu} d_{\mu, \bm k_2}&= W_{\bm k_1;\mu\nu} d_{\nu, \bm k_1}.
 \end{align}
In matrix notation, we perform the following steps to cast it into Hermitian form
  \begin{align}
 \notag (\Gamma^{-1}) \bar{d}&= W \bar{d}\\
 \notag U\Lambda U^\dag \bar{d}&= W \bar{d}\\
 \notag  \Lambda \bar{f}&= U^\dag W U \bar{f}\\
\bar{h}&= \Lambda^{-\frac{1}{2}} U^\dag W U  \Lambda^{-\frac{1}{2}} \bar{h}.
 \end{align}
Since $\Gamma_{\bm k_1,\bm k_2;\mu\nu}$ is spin-diagonal in our modelling, then so too are $U$ and $\Lambda^{-\frac{1}{2}}$. Moreover,  $\Lambda$ and $\Lambda^{-\frac{1}{2}}$ are completely diagonal, while $U=(U^{\mu\mu}_{\bm k,\chi_i})$ contains, as column vectors, the orthonormal basis of spatial harmonics of $\Gamma$, which we index by $\chi_i$. Using indices,
\begin{align}
 \label{h_gapeqn}
  h_{\chi_i}^\mu &=(\Lambda^{-\frac{1}{2}} )_{\chi_i,\chi_i}^{\mu\mu}  (U^\dag_{\chi_i, \bm k_1})^{\mu\mu} W_{\bm k_1}^{\mu\nu}  U_{\bm k_1, \chi_{j}}^{\nu\nu} (\Lambda^{-\frac{1}{2}} )_{\chi_{j},\chi_j}^{\nu\nu}  h_{\chi_j}^{\nu}.
 \end{align}
We solve for $h_{\chi_i}^{\mu}$ as eigenvectors and subsequently obtain the order parameter components $d_{\mu,\bm k}$ via the inverse transformation 
 \begin{align}
d_{\mu,\bm k}= U_{\bm k, \chi_{j}}^{\mu\nu} (\Lambda^{-\frac{1}{2}} )_{\chi_{j},\chi_j}^{\nu\nu}  h_{\chi_j}^{\nu}.
 \end{align}
 It is these $d_{\mu, \bm k}$ that are presented in the main text.

 Finally, we model the momentum-dependent factor $F_{\bm k_1, \bm k_2}$, appearing in the potential \eqref{Gamma_eqns}, as Lorentzian-shaped about the two partially filled bands $\varepsilon_{+,n,\bm k}$, with $n=1,2$, and with an exponential factor favoring small angle scattering, i.e. $\bm k_1\approx \bm k_2$,
 \begin{align}
 F_{\bm k_1, \bm k_2} &=  (1+a_0 e^{-a_1 |\bm k_1 - \bm k_2|}) L_{\bm k_1,\varepsilon_F} L_{\bm k_2,\varepsilon_F},\\
 L_{\bm k,\varepsilon_F}&= \frac{1}{1+a_2 \left[(\varepsilon_{+,1,\bm k}-\varepsilon_F)(\varepsilon_{+,2,\bm k}-\varepsilon_F)\right]^2/\varepsilon_F^4}.
 \end{align}
 The parameters $\{a_0,a_1,a_2\}=\{3, 30, 9000\}10^3$ are chosen to demonstrate the momentum-dependent admixture of order parameters. We work in units where the monolayer graphene lattice constant $a=1$.

\section{Perturbative treatment of the gap equation} \label{PerturbationTheory}
Sec. \ref{SC_minimal} presented the key characteristics of the superconducting order parameters $d_{\mu, \vec  k}$, based on a perturbative treatment; in this appendix we present the details of the perturbation expansion. For ease we reprint the gap equation (\ref{triplet_gapeqn}) of the complementary model of Sec. \ref{SC_minimal},
\begin{align}
\notag &d_{0,\vec k}= \sum_{\vec k}  \Gamma_{\bm k, \bm k'}^0  \left[{\mathbb V}^+_{T,\vec k'}d_{0,\vec k'} + {\mathbb V}^-_{T,\vec k'}\hat{\vec g}_{\vec k'}\cdot \vec d_{\vec k'}\right],\\
\label{Supp_gapeqn_full}
 &\vec d_{\vec k}=\sum_{\vec k} \Gamma_{\bm k, \bm k}^0 \bigg[{\mathbb V}^+_{T,\vec k'}\vec d_{\vec k'}+ {\mathbb V}^-_{T,\vec k'}\hat{\vec{g}}_{\vec k'}d_{0,\vec k'} + 2 {\mathbb V}^0_{T,\vec k'}[\hat{\vec g}_{\vec k'}(\hat{\vec g}_{\vec k'}\cdot\vec d_{\vec k'})-\vec d_{\vec k'}] \bigg].
\end{align}
We will treat terms $\propto  {\mathbb V}^-_{T,\vec k}$ as perturbations and note that these terms are responsible for mixing singlet and triplet components. In more compact notation, this is re-written as
  \begin{align}
\bar{d}&= \Gamma W \bar{d}.
 \end{align}
Our perturbation theory begins by splitting the matrix  $W=W^0 + W^1$, where $W^0 \propto \delta_{\mu\nu}$ and  the perturbation $(W^1)^{\mu\nu}$ is off-diagonal in spin; it contains SOC, via the $g_{\bm k,\mu}$-vectors as well as factors of ${\mathbb V}^-_{\bm k}$, which are the difference of thermal occupation factors of the spin-split bands and which changes sign in between the corresponding Fermi surfaces.

\vspace{1em}\noindent
{\bf $\bm A$-state with admixed triplet.} 
Consider a zeroth-order, spin-singlet ($\mu=0$) eigenstate, which necessarily belongs to the trivial spatial irrep $\chi_i \in A$. We denote this zeroth-order state as
\begin{align}
\bar{h}^{A} = (h^A_{0,\chi_i}, \bm 0),
\end{align}
which is written as a four-vector in spin, and with the singlet component, $h^A_{0,\chi_i}$, indexed by the harmonics $\chi_i$. The corresponding eigenvalue is denoted $E_A$. Hence we use the label $A$ to denote the linear combination of all $\chi_i \in A$, which would be found via direct diagonalisation of  \eqref{h_gapeqn} with $W\to W^0$.  Next, we denote a generic eigenstate $\bar{h}^{R}$ in a {\it spatial} irrep $R$ and with corresponding eigenvalue $E_R$. Note: (i) there can be multiple orthogonal eigenstates belonging to the same  irrep $R$, and (ii) the combination of spin and space may have a different combined spin-spatial irrep, but for the presentation here, it is convenient to talk about pure spatial irreps. 

Now we consider the perturbation $W^1$, and determine the first order correction to $\bar{h}^{A}$, i.e. find the $\bar{h}^{R}$ which become admixed via the perturbation. This is done via standard quantum mechanical perturbation theory, 
\begin{align}
\label{h_perturb_mat}
\tilde{\bar{h}}^{A} &= \bar{h}^{A} +  \sum_{R}\frac{1}{E_A-E_R} \left[(\bar{h}^R)^\dag \Lambda^{-\frac{1}{2}} U^\dag W^1U \Lambda^{-\frac{1}{2}} \bar{h}^{A}\right] \ \bar{h}^{R},
\end{align}
whereby the perturbing potential is $\Lambda^{-\frac{1}{2}} U^\dag W^1U \Lambda^{-\frac{1}{2}}$. 
Next, we can greatly simplify the expression by returning to the original $\bar{d}$ variables;  the transformation is $\bar{d}_R= U \Lambda^{-\frac{1}{2}} \bar{h}_{R}$, and here the $\bar{d}_R$ are vectors in the spatial irrep $R$. Applying this transformation to the perturbed-eigenvector expression \eqref{h_perturb_mat},
\begin{align}
U \Lambda^{-\frac{1}{2}} \tilde{\bar{h}}^\mu_{A} &= U \Lambda^{-\frac{1}{2}} \bar{h}^0_{A} + \sum_{R}\frac{1}{E_A-E_R} \left[\bar{h}_R^\dag \Lambda^{-\frac{1}{2}} U^\dag W^1U \Lambda^{-\frac{1}{2}} \bar{h}_{A}\right] \ U \Lambda^{-\frac{1}{2}} \bar{h}_{R},
\end{align}
we obtain
\begin{align}
\label{d_perturb_mat}
\tilde{\bar{d}}_{A} &= \bar{d}_{A} + \sum_R \frac{1}{E_A-E_R}  \left[\bar{d}_R{}^\dag  W^1 \bar{d}_{A}\right] \ \bar{d}_{R}.
\end{align}
Re-instating indices and the explicit form of $W^1$, which follows from the gap equation \eqref{Supp_gapeqn_full}, we arrive at 
\begin{align}
\label{d_perturb}
\notag \tilde{d}^A_{\mu,\bm k}  &= d^A_{0,\bm k}\delta_{\mu,0} +  \sum_{R, \bm k_1}\frac{1}{E_A-E_R}  \left[(d^R_{\mu,\bm k_1})^\dag  (W^1_{\bm k_1})^{\mu0} d_{0,\bm k_1}^A\right]  d^R_{\mu, \bm k}\\
&= d^A_{0,\bm k}\delta_{\mu,0} +  \sum_{R, \bm k_1}\frac{1}{E_A-E_R}  \left[(d^R_{\mu,\bm k_1})^\dag  {\mathbb V}^{-}_{\bm k_1} \bm g_{\bm k_1}^{\mu} d^A_{0,\bm k_1}\right]   d^R_{\mu, \bm k}.
\end{align}

The expression \eqref{d_perturb} represents the explicit result of our perturbation theory, without any approximations. Now, based on \eqref{d_perturb}, we introduce an approximate expression. First, examining the matrix element, $\sum_{\bm k_1}\left[(d^R_{\mu,\bm k_1})^\dag  {\mathbb V}^{-}_{\bm k_1} \bm g_{\bm k_1}^{\mu} d^A_{0,\bm k_1}\right]$, 
we see that it is nonzero only for those $d^R_{\mu,\bm k_1}$ that transform in the same irrep as ${\mathbb V}^{-}_{\bm k_1}\bm g_{\bm k_1}^{\mu} d_{\bm k_1, A}^0$. Since ${\mathbb V}^{-}_{\bm k_1}$ changes sign radially between the two spin-split Fermi surfaces, while the $g_{\bm k_1}^{\mu}$ change sign upon winding about the Fermi surfaces, then only those $d^R_{\mu,\bm k_1}$ that exhibit both the radial and angular sign changes will have an appreciable overlap. Therefore, maximal overlap occurs for $d^{R}_{\mu,\bm k; \text{projected}} \approx  {\mathbb V}^{-}_{\bm k} \bm g_{\bm k}^{\mu} d_{0,\bm k}^A$, and we arrive at the approximation 
\begin{align}
\label{dA_approx}
\tilde{d}^\mu_{\bm k, A}  \approx d^A_{0,\bm k}\delta_{\mu,0} + \beta \  {\mathbb V}^{-}_{\bm k} \bm g_{\bm k_1}^{\mu}  d_{\bm k, A}^0,
\end{align}
with numerical factors absorbed into the constant $\beta$. Upon reinstating the $\eta$-index and employing compact vector notation, the perturbed singlet $A$-state becomes
\begin{align}
\tilde{d}^A_{\bm k, \eta} & =  (d^A_{0,\eta\cdot\bm k}; \beta \ \eta {\mathbb V}^-_{T,\eta\cdot\vec k} \ \hat{\vec{g}}_{\eta\cdot\vec k} d^A_{0,\eta\cdot\bm k}),
\end{align}
Noting that ${\mathbb V}^{-}_{\bm k}$ changes sign between the fermi surfaces, while the $\bm g_{\bm k_1}^{\mu}$ winds about the fermi surface, together this gives a M\"obius-like spin-triplet texture to the admixed state, c.f. \cite{MobiusPRL}.  \\

\noindent 
{\it Comment.} It appears that from \eqref{d_perturb_mat}, the form of the potential $\Gamma$ does not show up and hence the admixture is set purely by the perturbation $W^1$. First, we can only arrive at \eqref{d_perturb_mat} if $\Gamma$ is invertible and positive definite. Second,  it is the $\bar{h}$-eigenvectors of \eqref{h_gapeqn} that are normalized, while the $\bar{d}$-vectors are found via the non-unitary transformation $\bar{d}= \Lambda^{-\frac{1}{2}} U \bar{h}$. Due to this non-unitary transformation, the potential $\Gamma$ then influences the magnitude of the $\bar{d}$-vectors.

\vspace{1em}\noindent
{\bf $\bm E$-state with admixed singlet.} 
Focusing now on the in-plane triplet eigenstate, which at zeroth-order is denoted $d^E_{j,\bm k} \in E$ IR, where index $j=1,2$ enumerates the in-plane triplet components, and has corresponding eigenvalue $E_{E}$. A first order perturbation in $\hat{W}^1$ admixes a singlet component into this triplet, 
\begin{align}
\label{d_perturb}
\tilde{d}^E_{\mu,\bm k}  &= d^E_{\mu,\bm k}(\delta_{\mu,1}+\delta_{\mu,2}) +  \delta_{\mu,0}\sum_{R, \bm k_1,j'}\frac{1}{E_{E}-E_{R}}  \left[(d^R_{0,\bm k_1})^\dag  (W^1_{\bm k_1})^{0{j'}} d^E_{j',\bm k_1}\right]  d^R_{0,\bm k}\\
&= d^E_{\mu,\bm k}(\delta_{\mu,1}+\delta_{\mu,2}) +  \delta_{\mu,0}\sum_{R, \bm k_1}\frac{1}{E_{E}-E_{R}}  \left[(d^R_{0,\bm k_1})^\dag  {\mathbb V}^{-}_{\bm k_1} \bm g_{\bm k_1} \cdot \bm d^E_{\bm k_1}\right]  d^R_{0,\bm k}
\end{align}
Noting that each component of $d^E_{j,\bm k}$ transforms trivially under a pure spatial rotation, then the $d^R_{0,\bm k}$ which posses a non-negligible overlap with ${\mathbb V}^{-}_{\bm k} \bm g_{\bm k} \cdot \bm d^E_{\bm k}$ must exhibit both the radial and angular sign changes. Like our discussion above, this leads to the approximate expression,
\begin{align}
\label{dE_approx}
\tilde{d}^E_{\mu,\bm k}  &\approx d^E_{\mu,\bm k}(\delta_{\mu,1}+\delta_{\mu,2})  +  \beta \ {\mathbb V}^{-}_{\bm k} \bm g_{\bm k} \cdot \bm d^E_{\bm k}\delta_{\mu,0}.
\end{align}
Here $\beta$ represents a numerical constant, different from \eqref{dA_approx}.
Re-instating the valley index $\eta$, and conforming to the more compact notation of the main text, this becomes,
\begin{align}
\tilde{d}^E_{\bm k} =(\beta  {\mathbb V}^{-}_{\eta\cdot\bm k} \bm g_{\eta\cdot\bm k}\cdot \bm d^E_{\eta\cdot\bm k}; \eta \bm d^E_{\eta\cdot\bm k}).
\end{align}
This perturbative expression shows that the admixed singlet exhibits sign changes going around the Fermi surfaces, due to $\hat{\vec{g}}_{\eta\cdot\vec k}\cdot {\bm d_{\eta\cdot\bm k}^E}$, and changes sign between the two spin-split Fermi surfaces, due to ${\mathbb V}^-_{T,\eta\cdot\vec k}$. Both such features are seen in the admixed singlet component $d_{0, \vec k}$ obtained from the full numerical solution of the gap equation.

\end{appendix}

\end{document}